\documentclass[aps,prl,twocolumn,amsmath,amssymb,longbibliography,superscriptaddress,floatfix]{revtex4-2}
\usepackage{mathtools}
\usepackage{amsmath,amssymb,amsfonts,bm}
\usepackage{graphicx}
\usepackage{xcolor}
\usepackage{bbold}
\usepackage{graphicx}
\usepackage{dcolumn}
\usepackage{epstopdf}
\usepackage{epsfig}
\usepackage[caption=false]{subfig}
\usepackage{url}
\usepackage{hyperref}
\usepackage{verbatim}
\usepackage[export]{adjustbox}
\usepackage{scalerel}
\usepackage{braket}
\usepackage{mathrsfs}
\usepackage{ulem}

\usepackage{amsthm}
\usepackage{enumitem}

\usepackage{multirow}
\usepackage{booktabs}
\usepackage{pifont}

\usepackage{colortbl}

\newcommand{\riesz}{\Omega}

\def\iii{j}

\def\ide{\mathbb{I}}

\newcommand{\tr}{\operatorname{tr}}
\newcommand{\Tr}{\operatorname{Tr}}

\def\Swap{\mathbb{S}}

\newcommand{\1}{\mathrm{I}}
\newcommand{\be}{\begin{equation}}
\newcommand{\ee}{\end{equation}}
\newcommand{\ba}{\begin{aligned}}
\newcommand{\ea}{\end{aligned}}

\newcommand{\bmult}{\begin{multline}}
\newcommand{\emult}{\end{multline}}

\newcommand{\lket}[1]{\mathinner{|#1\rangle\!\rangle}}
\newcommand{\lbra}[1]{\mathinner{\langle\!\langle#1|}}

\def\tind{\tau}

\DeclareMathOperator\arctanh{arctanh}

\graphicspath{{figures/}}

\renewcommand{\imath}{\mathrm{i}}

\newcommand{\aveE}[1]{\operatorname{\mathbb{E}}[#1]}

\newcommand{\aveEG}[1]{\operatorname{\mathbb{E}_G}[#1]}

\def\trho1{\tilde{\varrho}}

\begin{document}

\newcommand{\titleinfo}{Elusive phase transition in the replica limit of monitored systems}
\title{\titleinfo}

\author{Guido Giachetti}
\affiliation{Laboratoire de Physique de l'\'Ecole Normale Sup\'erieure, CNRS, ENS \& PSL University, Sorbonne Universit\'e, Universit\'e Paris Cit\'e, 75005 Paris, France}

\author{Andrea De Luca}
\affiliation{Laboratoire de Physique de l'\'Ecole Normale Sup\'erieure, CNRS, ENS \& PSL University, Sorbonne Universit\'e, Universit\'e Paris Cit\'e, 75005 Paris, France}

\date{\today}

\begin{abstract}
\noindent
We study an exactly solvable model of monitored dynamics in a system of $N$ spin-$1/2$ particles with pairwise all-to-all noisy interactions, where each spin is continuously weakly measured along a random direction. Using the replica trick to incorporate the Born-rule weighting of measurement outcomes, we obtain an exact large-$N$ description of purification and of the statistics of local observables. We find that the nature of the phase transition strongly depends on the number $n$ of replicas: non-perturbative logarithmic corrections appear in the physically relevant $n\to1$ limit and destroy the purifying phase present at finite integer $n$. As a consequence, the purification time of an initially mixed state is always exponentially long in the system size, even at arbitrarily large measurement rate.
\end{abstract}

\maketitle


\paragraph{Introduction. ---}

The out-of-equilibrium dynamics of many-body quantum systems has attracted great attention in recent years because of open questions related to chaos, thermalisation, and ergodicity~\cite{dalessio2016,RevModPhys.91.021001}, together with the opportunity to engineer novel phases and quantum technologies~\cite{Blatt2012, Schafer2020}. A particularly exciting recent development is the study of individual trajectories in noisy~\cite{PhysRevX.13.011043, PhysRevX.13.011045, Bernard_2021, PhysRevX.8.021014, PhysRevX.7.031016, PhysRevB.99.174205} and monitored systems~\cite{PhysRevA.36.5543, PhysRevA.58.1699, Gisin_1992}, where the unitary dynamics due to internal interactions competes with quantum measurements. Protocols of this kind have several applications, ranging from control via feedback to quantum computation~\cite{PhysRevA.62.062311, leung2004quantum, PhysRevLett.101.010501}, error correction~\cite{PhysRevLett.125.030505}, and purification~\cite{Ticozzi2014,Masanes2017, PhysRevX.10.041020}. In particular, collective spin systems under continuous monitoring---possibly combined with feedback control---have emerged as a paradigmatic setting where the measurement backaction qualitatively reshapes the many-body dynamics and its phase transitions~\cite{daley2014quantum, ashida2016quantum, ivanov2020feedback}.
Much attention has also been devoted to the dynamics of entanglement: while closed quantum dynamics generically leads to linear entanglement growth before a volume-law saturation~\cite{Kormos2017, PhysRevLett.109.017202, Calabrese_2005, PhysRevLett.111.127205, PhysRevX.7.031016, PhysRevX.8.041019, PhysRevX.9.021033}, quantum measurements drive a sharp transition, as a function of the measurement rate, between a weak-measurement volume-law phase and a strong-measurement area-law phase~\cite{PhysRevX.9.031009,PhysRevB.98.205136, PhysRevB.99.224307}.
Along with some solvable~\cite{PRXQuantum.2.010352, PhysRevLett.127.140601, bentsen2021measurement, PhysRevLett.128.010603} and numerically tractable cases~\cite{PhysRevB.100.134306, PhysRevB.104.155111, PhysRevB.101.104302, PhysRevLett.130.220404, PhysRevB.105.104306}, a rich literature has supported, through numerical studies, the existence of a second-order measurement-induced phase transition (MIPT), for a large number of interacting models and measurement protocols~\cite{PhysRevResearch.2.013022, PhysRevB.106.024305, PhysRevB.102.035119, PhysRevB.106.214316, fuji2020continuous, biella2021subradiance}. 
Because of their connections with entanglement properties of individual trajectories, MIPTs have also been interpreted as a change in the classical simulatability of the quantum dynamics~\cite{PhysRevLett.91.147902, PhysRevLett.93.040502, PhysRevLett.125.070606}. Gaussian quadratic models, such as free fermions, under monitoring deserve a special role~\cite{PhysRevA.107.032215, Fidkowski2021howdynamicalquantum}: in such a case, the dynamics of measurements can be treated numerically efficiently and the fragility of the volume-law phase for arbitrarily weak measurements has been pointed out~\cite{10.21468/SciPostPhys.7.2.024, Fidkowski2021howdynamicalquantum}, along with the possibility of more peculiar transitions between area and subvolume (e.g., logarithmic) scaling of entanglement~\cite{PhysRevLett.126.170602, PhysRevX.11.041004, PhysRevLett.128.010605, PhysRevB.105.094303, PhysRevResearch.4.033001, PhysRevResearch.2.023288, PhysRevResearch.4.043212, PhysRevB.105.064305, PhysRevB.105.L241114, PhysRevB.103.224210}.
\begin{figure}[t]
    \centering 
    \vspace{0.2cm}
    \includegraphics[width=0.72\columnwidth]{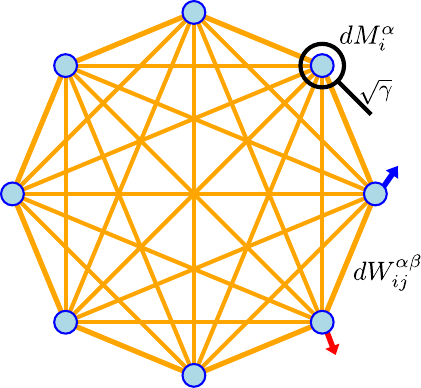}
    \caption{Schematic description of the model: $N$ spin-$1/2$ particles interact through fully connected isotropic noisy couplings $dW^{\alpha \beta}$. Each spin is also coupled to ancillary qubits $\ket{\mathcal{A}}$, which are projectively measured.}
    \label{fig:Protocol}
\end{figure}
A major difficulty in studying monitored systems is the nonlinear role of Born's rule, since the probability of a measurement record depends on the evolving quantum state. A standard way to deal with this nonlinearity is the replica trick~\cite{PhysRevB.100.134203,fava2023nonlinear}: one studies $n$ copies of the system with unbiased measurement noise and only at the end takes the analytic continuation $n\to 1$ that restores the physical weighting of trajectories. Attempts to construct a quantitative theory of MIPTs have mainly been based on this approach together with large-$N$ limits and Landau-Ginzburg-style field theories~\cite{nahum2023renormalization, PRXQuantum.2.010352, PhysRevB.102.064202}
with a conformally-invariant critical point~\cite{PhysRevLett.128.050602}. For free-fermion models, a nonlinear sigma model approach~\cite{doi:10.1080/00018730902850504, fava2023nonlinear} has shown that the renormalization-group flow in the replica limit can drastically alter the entanglement scaling~\cite{fava2023nonlinear} or even remove the transition altogether~\cite{poboiko2023theory}---subtleties that emerge only in the $n\to1$ limit and are inaccessible at integer $n>1$.

In this Letter, we introduce an exactly solvable model of monitored dynamics, composed of $N$ spin-$1/2$ particles with all-to-all noisy pairwise interactions. Each spin is subjected to weak continuous measurements along a random direction. We derive a mean-field theory in which the physics is encoded in a scalar order parameter and its effective free energy. 
We are also able to access the relevant $n\to1$ limit: our main result is that in this limit the free energy displays non-perturbative logarithmic corrections which stabilize the volume-law phase for arbitrary measurement strength.
This implies that the purification time of a mixed state is always exponentially long in the system size as the purifying phase present for finite $n$ disappears. We validate our theory by looking at the full statistics of single-site observables. 
\paragraph{The model. ---}
Let us consider the Hilbert space $\mathcal{H} = [\mathbb{C}^{2}]^{\otimes N}$ of $N$ spin $1/2$, subject to the isotropic noisy time-evolution, generated by the Hamiltonian increment
\begin{equation}
\label{eq:ham}
    d\hat H_0 = \frac{J}{\sqrt{N}} \sum_{1\leq i<j 
    \leq N} \sum_{\alpha,\beta} dW_{ij}^{\alpha\beta} \hat S_i^\alpha \hat S_j^{\beta}
\end{equation}
where $J$ is the coupling strength and $\hat S_i^{\alpha}$ ($\alpha = x,y,z$) are spin $1/2$ operators acting on the local Hilbert space at site $i$ and $dW_{ij}^{\alpha \beta}$ are centered Gaussian white noises, with $\aveE{dW_{ij}^{\alpha \beta} dW_{i'j'}^{\alpha' \beta'}} = dt \delta^{\alpha \beta, \alpha' \beta'}_{i'j',ij}$. The increment of the density matrix due to the unitary dynamics thus reads (in Ito's convention)
\begin{equation}
\label{eq:rhouni}
[d\rho]_{\rm uni} 
= -\imath [d\hat H, \rho] + \frac{J^2}{N} \sum_{1\leq i<j 
    \leq N}\sum_{\alpha,\beta} \mathcal{D}_{S_i^\alpha S_j^\beta}[\rho]
\; ,
\end{equation}
where we introduced the dephasing superoperator $\mathcal{D}_{\hat O}[\rho] = - \frac12 [\hat O, [\hat O, \rho]]$. On top of this unitary evolution, the $N$ spins, initially prepared in the maximally-mixed state $\rho_0 = \ide/2^N$ are continuously monitored in the standard framework of homodyne detection~\cite{PhysRevA.36.5543, PhysRevA.58.1699, Gisin_1992} for $t\in[0,T]$. A convenient microscopic realization is to discretize time $t = \tau \Delta t$, $\tau \in \mathbb{N}$, and couple, for each $\Delta t$ interval the component $\alpha$ of spin $i$ to an ancilla qubit. The ancilla is then projectively measured along the $z$ direction leading to an outcome $a_i^\alpha(\tau = t/\Delta t) = \pm 1$~\cite{SM}. This construction has a well-defined $\Delta t\to0$ limit, where the outcomes $\mathbf{a} = \{a_i^\alpha(\tau)\}_{i,\alpha,\tau}$, with $\tau = 1,\ldots, \mathsf{T} = T/\Delta t$, are organized as measurement records $M_i^\alpha(t) := \Delta t^{-1/2} \sum_{\tau'\Delta t <  t} a_i^\alpha(\tau')$ that converge to well-defined stochastic processes. Here, we focus on isotropic and homogeneous case where all components and all spins are measured with a single rate $\gamma$. Note that, despite its fully-connected structure, the model differs in an essential way from dissipative or monitored Lipkin-Meshkov-Glick--type models~\cite{lipkin1965validity, morrison2008dynamical, santini2025semiclassical}: since the couplings $dW_{ij}^{\alpha\beta}$ are independent noises for each pair of sites and of spin components, the total spin is not conserved and the dynamics cannot be reduced to that of a single collective spin.
The corresponding stochastic Schr\"odinger equation (SSE) for the many-body density matrix $\rho$ (see Sec. S1 in \cite{SM}) is inherently non-linear because of the measurement process. To avoid this difficulty, it is more convenient to introduce the non-normalized operator
$\tilde{\rho}_{\mathbf{a}}(t)= \mathsf{K}_{\mathbf{a}}\rho_0 \mathsf{K}_{\mathbf{a}}^\dagger$, with $\mathsf{K}_{\mathbf{a}}$ the Kraus operators ($\sum_{\mathsf{a}} \mathsf{K}_{\mathbf{a}}^\dagger \mathsf{K}_{\mathbf{a}} = \ide$) of the measurement sequence $\mathbf{a}$~\cite{nielsen2002quantum}. In the limit $\Delta t \to 0$, $2^{\mathsf{T}/2} \mathsf{K}_{\mathbf{a}} \to \mathcal{K}_{M}$, the non-unitary stochastic evolution operator generated by the infinitesimal increment
\begin{equation}
\label{eq:hamfull}
    d\hat H = 
   d\hat H_0+ \imath \sqrt{\gamma}\sum_{\alpha,i} dM_i^\alpha \hat S_i^\alpha - \imath \gamma \sum_{i} \hat S^{2}_i dt \;,
\end{equation}
so that
$\mathcal{K}_M(t+dt)=e^{-\imath d\hat H}\mathcal{K}_M(t)$
and
$\tilde\rho_M+d\tilde\rho_M = e^{-\imath d\hat H} \tilde \rho_M e^{\imath d\hat H^\dag}$.
Consider now any functional $F[\rho]$ of the trajectory state. According to the Born rule, $\tr \tilde{\rho}_M$ gives the probability of a given realization of the noise. As a consequence,
\begin{equation}
\label{eq:Frhoreplica}
    \aveE{F[\rho]}
    =
    \mathbb{E}_G \left[ F\!\left[\tilde{\rho}_M/ \tr \tilde{\rho}_M\right] \tr \tilde{\rho}_M \right].
\end{equation}
The left-hand side averages over SSE realizations, which automatically enforce normalization and Born-rule weighting; on the right-hand side the Born probability is carried explicitly by $\tr\tilde\rho_M$, so the $dM_i^\alpha$ in Eq.~\eqref{eq:hamfull} are \textit{unbiased} Gaussian noises, $\aveEG{dM_i^\alpha dM_j^\beta} = dt \delta_{ij} \delta^{\alpha\beta}$.

For linear functionals, the averaged state $\aveE{\rho} \equiv \mathbb{E}_G[\tilde{\rho}]$ follows a linear Lindblad evolution and generically relaxes to the structureless infinite-temperature state independently of the measurement rate $\gamma>0$. Thus, we focus on nonlinear functionals, choosing in particular $F[\rho] = \Tr[\Omega_k \rho^{\otimes k}]$ for an appropriate choice of $\Omega_k$, i.e. multilinear on a finite number $k$ of copies of the density matrix. This includes higher spectral moments and moments of local expectation values; for instance, the purity $\Pi = \Tr[\rho^2]$ corresponds to $k=2$ and $\Omega_2= \Swap$, the swap operator exchanging the two replicas. The replica trick is then used to avoid the denominator in Eq.~\eqref{eq:Frhoreplica}: by embedding $\riesz^{(n)}=\Omega_k\otimes\ide^{\otimes(n-k)}$, so that
\begin{equation}
\label{eq:gaussE}
    \mathbb{E} [F[\rho]] \equiv  \lim_{n\to 1}\mathbb{E}_G[\Tr^{(n)}[\riesz^{(n)} \tilde\rho^{\otimes n}_M ]] \ .
\end{equation}
where the superscript denotes the trace on the $n$-replicas Hilbert space. In the following, we focus on observables that can be expressed in this form, where $\Omega^{(n)} = \otimes_{i=1}^N \Omega^{(n)}_i$ has a tensor-product structure over all sites.
\begin{figure}
    \centering 
    \vspace{0.2cm}
    \includegraphics[width=0.99\columnwidth]{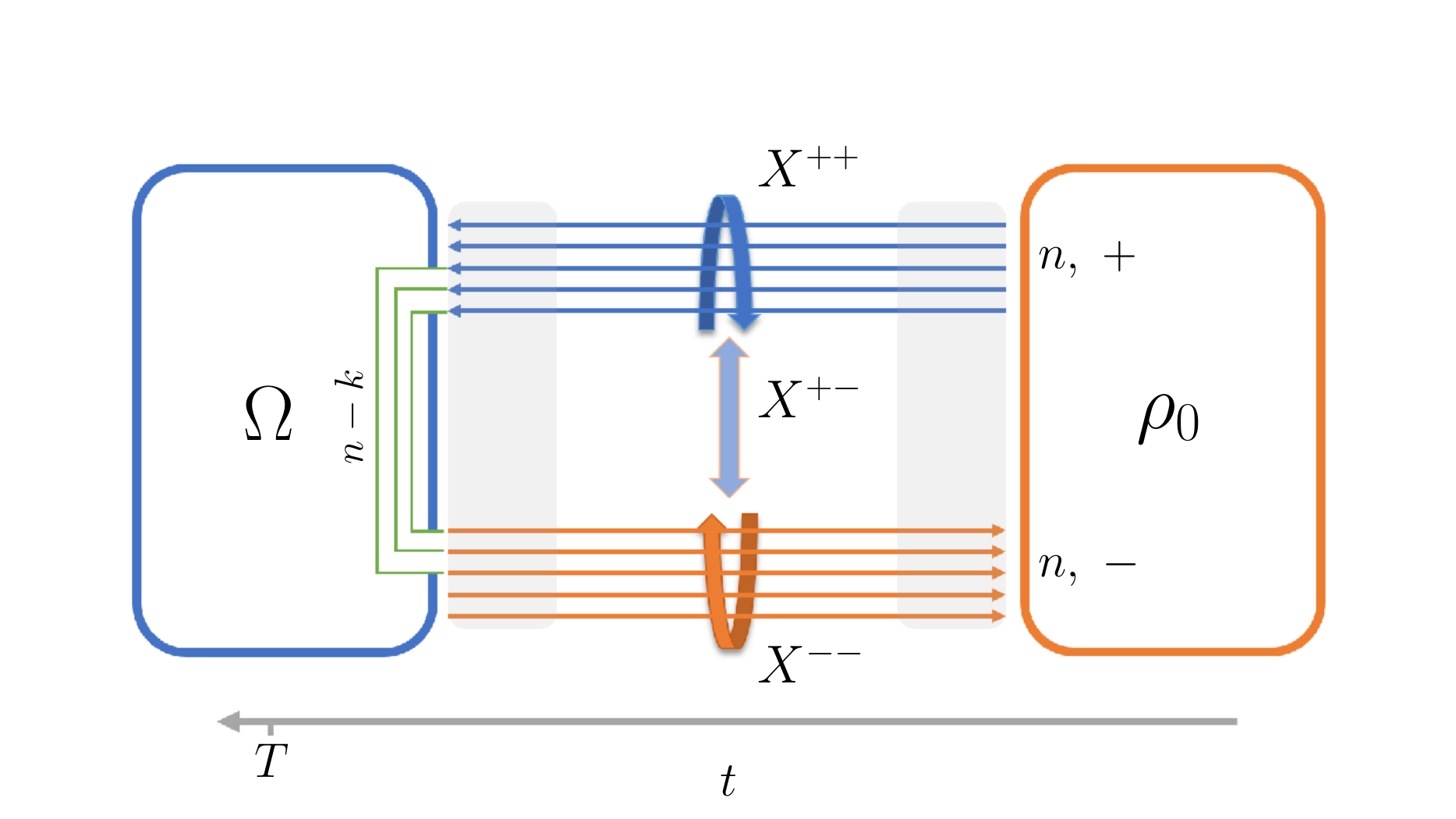}
    \caption{Schematic description of the Keldysh path integral that leads to Eq.\,\eqref{eq:Idefmain}. After averaging over the unitary noise and the effect of measurements, one introduces Hubbard-Stratonovich fields $X^{\sigma\sigma'}_{ab}(t)$ coupling replicas $a,b$ on contours $\sigma,\sigma'=\pm$. In the $\mathcal C$-invariant sector relevant below, the same-contour fields are identified, $X^{++}(t)=X^{--}(t)$, while the mixed-contour fields satisfy $X^{-+}_{ab}(t)=X^{+-}_{ab}(t)$. If $\riesz^{(n)}$ does not explicitly break the replica symmetry, the bulk fields are time independent apart from the shaded transient region. The operator $\riesz^{(n)}$ acts trivially on $n-k$ replicas.}
    \label{fig:Keldysh}
\end{figure}

\paragraph{Replica path integral. ---}
It is useful to express the expectation value in Eq.~\eqref{eq:gaussE} by means of a Keldysh spin-coherent path integral~\cite{kamenev2023field, fradkin2013field} (see Fig.~\ref{fig:Keldysh}). This amounts to introducing for each site $i=1,\ldots, N$, replica $a=1,\ldots,n$ and Keldysh contour $\sigma = \pm$ a time-dependent classical vector field $\mathbf{s}_{i,a,\sigma}(t)$. 
As shown in~\cite{SM}, due to the all-to-all nature of the couplings, one can introduce Hubbard-Stratonovich fields $X_{ab}^{\sigma\sigma'}(t)$ which allow reducing the path-integral to an effective single-site $(0+1)$ dimensional theory~\cite{A_J_Bray_1980}. 
More formally, for the replicated expectation value in Eq.~\eqref{eq:gaussE}, one finds $\mathbb E_G[\Tr^{(n)}[\riesz^{(n)} \tilde\rho^{\otimes n}_M ]]
    \sim
    \int \mathcal{D}[X] \exp(-NT \mathcal{I}_n[X])$, where $\mathcal{D}[X]$ denotes the product measure over the independent components of the fields $X^{\sigma\sigma'}_{ab}(t)$ (see~\cite{SM}), with the effective action
\begin{multline}
    \mathcal{I}_n[X]
    =-\int^T_0 \frac{dt}{4T} \sum_{ab,\sigma,\sigma'}
    \sigma\sigma' \bigl(X^{\sigma\sigma'}_{ab}(t)\bigr)^2\\
    - \frac{1}{T} \ln \Tr \Bigl[\riesz^{(n)} \trho1^{(n)}(T)\Bigr] .
\label{eq:Idefmain}
\end{multline}
Here the fields $X^{\sigma \sigma'}_{ab}(t)$ are conjugate to the replicated spin bilinears $\mathbf{s}_{\sigma,a} \cdot \mathbf{s}_{\sigma',b}$ and $\trho1^{(n)}(T)$ is the effective \textit{single-site} (unnormalised) density matrix for $n$ replicas, obtained by evolving $\trho1^{(n)}(0) = \mathbb{1}/2$ with $ \frac{d}{dt} \trho1^{(n)}
    = \mathcal{L}_X^{(n)}(\trho1^{(n)})$ and
\begin{multline}
\label{main:mathLn}
    \mathcal{L}_X^{(n)}(\trho1^{(n)}) :=
    \frac{1}{2}\sum_{ab,\sigma,\sigma',\alpha}
    \bigl(\gamma-\sigma\sigma' X^{\sigma\sigma'}_{ab}(t)\bigr)
    \hat S^\alpha_{a,\sigma} \hat S^\alpha_{b,\sigma'}[\trho1^{(n)}]\\
    -\frac{3}{2}\gamma n \trho1^{(n)} ,
\end{multline}
where $\hat S^\alpha_{a,+}[\rho]\equiv \hat S^\alpha_a\rho$ and $\hat S^\alpha_{a,-}[\rho]\equiv \rho \hat S^\alpha_a$ are compact notation for left/right multiplication. The superoperator $\mathcal{L}_X^{(n)}$ depends parametrically on the time-dependent saddle fields $X_{ab}^{\sigma\sigma'}(t)$. In the $N \rightarrow \infty$ limit the path integral is dominated by saddle points of $\mathcal{I}_n[X]$. The corresponding self-consistency equations, written explicitly in Appendix~B [see Eqs.~\eqref{app:saddleX-general} and \eqref{app:saddleqrX}], express the Hubbard-Stratonovich fields in terms of replicated single-site expectation values computed with the effective evolution \eqref{main:mathLn}.

As discussed in \cite{SM}, the action \eqref{eq:Idefmain} has a
$(S_n \times S_n)\rtimes \mathbb{Z}_2$ symmetry.
For $(P_+, P_-) \in S_n \times S_n$, one can independently relabel the $+$ and $-$ replicas, i.e. $X_{ab}^{\sigma, \sigma'} \to X_{P_{\sigma(a)}, P_{\sigma'(b)}}^{\sigma, \sigma'}$. The $\mathbb Z_2$ exchanges the $\pm$ contours, $X_{ab}^{\sigma,\sigma'} \to (X_{ab}^{-\sigma, -\sigma'})^\ast$, where the star denotes the complex conjugate (see Appendix~B and the Supplemental Material for the explicit realization). Numerical checks at $n=2,3$ indicate that this $\mathbb Z_2$ remains unbroken with a real field $X$, so in the branch relevant here, we take $X^{++}=X^{--}$ and $X^{-+}_{ab}=X^{+-}_{ab}$. At small enough $\gamma$, a symmetry breaking can occur where the two permutation symmetries lock to each other only up to a relative permutation. Each broken-symmetry sector is then labeled by $P\in S_n$ and is characterized by the residual symmetry $P_- \in S_n$ with
\begin{equation}
\label{eq:symmbreakP}
    P_+=P\,P_-\,P^{-1}.
\end{equation}
and $n!$ degenerate vacua in correspondence of $P$. This symmetry breaking is directly tied to slow purification\cite{PhysRevX.10.041020,bentsen2021measurement}: $\trho1^{(n)}(0)$ lies in the identity sector $P=\mathbb 1$, while $\riesz^{(n)}$ selects the sector of the transposition $P = (1,2)$. The relevant saddle must therefore contain an instanton interpolating between these two vacua, leading to $\mathbb{E}[\Pi] \sim T e^{-NT \mathcal{I}^\ast}$, where $\mathcal I^\ast$ is the corresponding intensive free-energy barrier in the $n\to1$ limit.

\paragraph{Saddle-point solutions --} 
We now investigate the possibility of a symmetry breaking of the type \eqref{eq:symmbreakP} for sufficiently small $\gamma > 0$. To solve the saddle-point equations for $\mathcal{I}_n[X]$ for arbitrary $n$, we consider the Ansatz
\begin{equation}
\label{eq:replicaansatz}
    \bar{X}^{++}_{ab} = q(t) + \Bigl(\frac3 4 - q(t) \Bigr)\delta_{ab} , \; \bar{X}^{+-}_{ab} = r(t) + X(t) \delta_{ab} .
\end{equation} 
where $r(t),q(t), X(t) \in \mathbb{R}$. For $X(t) = 0$, Eq.~\eqref{eq:replicaansatz} provides the most general Ansatz which is completely invariant under the $S_n \times S_n \rtimes \mathbb{Z}_2$ symmetry described above. A $X(t) \neq 0$ signals a breaking of this symmetry according to Eq.~\eqref{eq:symmbreakP} with the particular choice $P = \mathbb{1}$. Note that the diagonal component is fixed by the saddle-point equation $X^{++}_{aa}=\langle \hat{\mathbf S}_a^2\rangle=3/4$ (see Appendix~B).
The saddle-point solution determines the fields $q(t), r(t), X(t)$ for every $t \in [0, T]$ via the replicated dynamics \eqref{main:mathLn} of a single spin. For large $T$ and $0 \ll t \ll T$, the bulk dynamics effectively become time-independent and $\mathcal{L}^{(n)}_X$ becomes a static generator. The replicated single-site problem arises from the collection of $n$ spins $1/2$, that we can decompose into irreducible representations of $SU(2)$. Let $\Pi_\ell$ be the orthonormal projectors onto the sector of fixed total spin $\mathbf S_{\rm tot}^2=\ell(\ell+1)$, with $\hat S_{\rm tot}^\alpha=\sum_a\hat S_a^\alpha$. Rotational invariance, together with the isotropy of the initial state $\tilde\varrho^{(n)}(0)=\mathbb{1}/2$, implies that the dynamics generated by $\mathcal{L}_X^{(n)}$ stays within $\operatorname{Span}[\{\hat\Pi_\ell\}]$, reducing the problem to the finite matrix $(\mathcal{L}_X^{(n)})_{\ell, \ell'} := \Tr[\Pi_{\ell'} \mathcal{L}_X^{(n)} \Pi_{\ell}]$ (see Appendix~B).
One parametrizes~\cite{SM}
\begin{equation}
\label{eq:Lprojll}
     (\mathcal L_X^{(n)})_{\ell,\ell'}
    =
    (2\gamma+r-q)\,L_{\ell,\ell'}
    + \frac{3n}{4}\left(q-2\gamma-\frac34\right)\delta_{\ell,\ell'},
\end{equation}
where the matrix $L_{\ell, \ell'} :=L_{\ell, \ell'}(x)$ is tridiagonal and only depends on the reduced order parameter $x = X/(2 \gamma + r - q)$~(see Eq.~\eqref{eq:Lmatrdef} for the explicit form). Solving the saddle point equation for  $q$ and $r$, one arrives at the reduced action, which reads (up to an inessential constant)
\begin{equation}  \label{main:mathIx}
\begin{split} 
    \mathcal{I}_n(x) = \frac{\left[ n(n+x)\left( \frac{3}{4} + 2 (n-1) \gamma \right) - \Lambda_n (x) \right]^2}{2n(n-1)(n+2x)} 
\end{split} \ . 
\end{equation}
where $\Lambda_n(x)$ is the largest eigenvalue of $L$. 
The cases $n=2$ and $n=3$ are already instructive: for $n=2$, in agreement with~\cite{bentsen2021measurement}, as $S_{n=2} \equiv \mathbb{Z}_2$, we find an Ising-like second-order transition at $\gamma_c = 1/4$; whereas for $n \geq 3$ the transition is discontinuous ($\gamma_c = \sqrt{2}/12 \approx 0.12$). The physically relevant limit $n\to 1$ can nevertheless be accessed exactly by expanding $\Lambda_n (x)$ asymptotically at large $x$, analytically continuing the coefficients to $n \to 1$, and resumming the resulting series\cite{martin2010exactly}.
As the preservation of the trace ensures that $\mathcal{I}_{n=1}(x) \equiv 0$,
we can set $\iii(x) \equiv \lim_{n\to1} \mathcal{I}_n(x)/(n-1)$ and 
find (see \cite{SM} for details) 
\begin{equation} \label{main:In=1}
    \iii(x) =\frac{1}{8(1+2x)} \left[ \left( 4\gamma + 1\right) (1+x)   - x k(x)\right]^2,
\end{equation}
where $k(x) = K_0(x)/K_1(x)$ and $K_p(x)$ denotes the order--$p$ modified Bessel function of the second kind. This asymptotic resummation leading to $\iii(x)$ is further justified in~\cite{SM} through a more direct analytical route, by expanding the three-term recurrence relation for the spectrum of $L$ around $n=1$, which closes exactly in terms of modified Bessel functions.
The function $\iii(x)$ has only one minimum at a finite $x^{*}(\gamma)>0$, regardless of the value of $\gamma>0$. By taking the thermodynamic limit first, and then the physical limit $n \rightarrow 1^{+}$, we thus find that the model does not exhibit any phase transition: rather, the symmetry is broken for every finite $\gamma$ (i.e., $\gamma_c = \infty$). This result shows a non-perturbative behavior emerging in the replica limit. Indeed the general Landau-Ginzburg picture, exposed in \cite{nahum2021measurement,nahum2023renormalization}, and consistent with the phenomenology of the $n=2$ case~\footnote{The expansion of the action for $n=3$ also takes this form but the transition is first-order due to the presence of a separate minimum, a feature therefore beyond the small $x$ expansion. This is somewhat reminiscent of what happens for the $q$-state Potts model in $2d$, displaying a second-order phase transition for $1\le q \le 4$ and a first-order one for $q>4$.} \cite{Duminil-Copin2017}, is based on a series expansion of the $n$-replica free-energy around $x=0$. 
Here however, for small $x$, $\iii(x) \sim x^2 \ln(x)$ 
suggesting the non-commutativity of the $x\to 0$ and $n\to1$ limits. We leave a field-theory justification of this phenomenology to further studies, but we suggest it might result from the degeneracy of the anomalous dimensions of different operators happening at $n \to 1$, a mechanism analogous to the emergence of logarithmic corrections in certain non-unitary conformal field theories~\cite{cardy1999logarithmic, Cardy_2013}. 
A similar technique can be employed in the \textit{forced measurement} $n\to 0$, where Born's probability is discarded and all measurement outcomes are equally weighted~\cite{PhysRevB.100.134203}. Again, no phase transition appears consistently with the general tendency to weaken the effect of measurements decreasing $n$.

\begin{figure}
    \centering 
    \includegraphics[width=0.99\columnwidth]{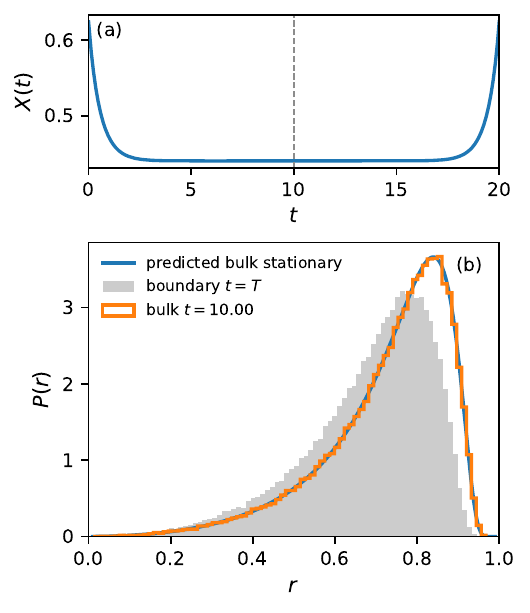}
    \caption{\textit{(a)} Behavior of $X(t)$ for $\gamma = 0.38$, obtained by solving numerically Eq.\,\eqref{main:drho2} together with the self-consistency condition and compared with the analytical bulk prediction. \textit{(b)} Histogram of the boundary distribution $\mathsf{r}(T)$ compared with the predicted bulk stationary distribution $P_{\rm stat}(\mathsf{r})$ for $\gamma = 0.38$. The difference between the two reflects the distinct role of the time boundary and of the stationary bulk regime.}
    \label{fig:panel}
\end{figure}

\begin{figure}
    \includegraphics[width=0.92\columnwidth]{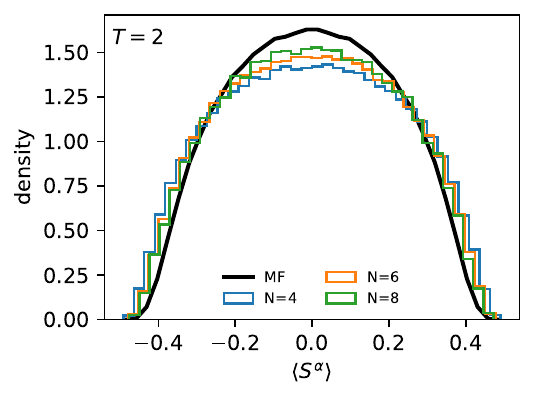}
    \par\smallskip
    \caption{Comparison between the self-consistent single-spin mean-field dynamics and the exact monitored evolution of the full density matrix for $N=4,6,8$ at $\gamma=0.38$. The plot shows the full distribution of one spin component at $T=2$. The mean-field prediction is obtained from the boundary samples of $\mathsf{r}(T)$ by isotropic projection, i.e. by assigning each sample $\mathsf{r}_j$ a uniform contribution on $[-\mathsf{r}_j/2,\mathsf{r}_j/2]$, while the exact data are averaged over all sites and spin components.}
    \label{fig:mfexactdist}
\end{figure}

\paragraph{Decoupling replicas. ---} 
The Ansatz \eqref{eq:replicaansatz} not only reduces the saddle problem to a scalar action, but also yields a direct single-spin unraveling of the physical limit $n\to1$. This framework gives access to the statistics of single-site observables, via the moments $\mathbb{E}[\Tr[\rho S_{i,\alpha}]^k]$, obtained in the replica construction by choosing on the $i$-th site $\riesz^{(n)}_i = (\hat S_{i}^{\alpha})^{\otimes k} \otimes \ide^{\otimes(n-k)}$ and the identity for sites $\neq i$.
Making again use of Eq.~\eqref{eq:Frhoreplica}, we can now treat the single-spin dynamics via a SSE, which reads
\begin{multline} \label{main:drho2}
    d \varrho = 
    - \text{i} \sqrt{\gamma_2(t)} dW^{\alpha} [\hat S^{\alpha} , \varrho] +  \sqrt{\gamma_1(t)} dY^{\alpha} \lbrace \delta \hat S^{\alpha}, \varrho \rbrace\\ 
    +  \left( \frac{3}{4} + \gamma \right) \mathcal{D}_{\hat S^\alpha}[\varrho]
\end{multline}
where $\delta\hat S^\alpha = \hat S^\alpha - \langle \hat S^\alpha\rangle$, $2 \gamma_1 (t) = 2 \gamma + 3/4 - X(t) - q(t)$, $2 \gamma_2 (t) = 3/4 - X(t) + q(t)$ and $dW^{\alpha}$, $dY^{\beta}$ are independent Wiener processes ($dW^{\alpha}dW^{\beta} =  dY^{\alpha} dY^{\beta} = \delta^{\alpha \beta} dt$) corresponding to effective single-spin unitary and measurement noise, respectively.
Note that we eliminated the parameter $r(t)$, as in the limit $n\to 1$, $r(t) + X(t) = 3/4$ (see $n\to1$ limit of~\eqref{app:saddleqrX}). For the remaining parameters $q(t), X(t)$, we can use that the bulk of sites $j \neq i$ in the identity sector $\Omega_j^{(n)} = \mathbb{1}$. There is therefore a time reflection symmetry, $t \to T-t$ , as $\varrho_0^{(n)} = \Omega_j^{(n)}/2$. It is useful to view the contribution from time intervals $[0,t]$ and $[t, T]$ as two distinct realisations $\varrho^{(1,2)}$ of the same stochastic dynamics Eq.~\eqref{main:drho2}. The self-consistency equation for $X(t)$ (and similarly for $q(t)$, see \cite{SM}) therefore takes the compact form
\begin{align} \label{main:selfconstX}
X(t) = \frac{3}{4} - 2 E\left(\frac{\sum_\alpha|\tr[\varrho^{(1)}(t) \hat S^\alpha \varrho^{(2)}(T-t)]| ^2}{\tr[\varrho^{(1)}(t) \varrho^{(2)}(T-t)]}\right)
\;.
\end{align}
Writing $\varrho(t)=\mathbb 1/2+\mathsf{r}_\alpha(t)S^\alpha$ and exploiting isotropy reduces the problem to a scalar stochastic dynamics for $\mathsf{r}(t)=\sqrt{\mathsf{r}_\alpha \mathsf{r}_\alpha}$. In practice, from an initial guess for $X(t), q(t)$, we generate a population $\{\mathsf{r}^{(\mu)}(t)\}_{\mu=1}^{N_{\rm sample}}$ for $t \in [0, T]$; we then average over $\mathsf{r}^{(1,2)}$ independently drawn from the sample to compute a new guess for $X(t), q(t)$. After few iterations, the procedure leads to  convergence to a fixed point for all $t\in[0,T]$, see Fig.~\ref{fig:panel}a. In the bulk $0\ll t\ll T$, the process reaches a stationary law $P_{\rm stat}(\mathsf{r})$, that can be determined analytically
\begin{equation} \label{main:Pstat}
    P_{\rm stat} (\mathsf{r}) = \frac{1}{Z} \frac{\mathsf{r}^2}{(1-\mathsf{r}^2)^3} e^{-2x(1-\mathsf{r}^2)^{-1}},
\end{equation}
which determines the bulk value of $X$ and reproduces the same stationarity condition $\iii'(x^\ast)=0$ obtained from Eq.~\eqref{main:In=1} (see also~\cite{SM}). This agreement provides an independent check of the analytic continuation.
For $x=0$, $P_{\rm stat}(\mathsf{r})\to\delta(\mathsf{r}-1)$ and the effective spin purifies, whereas any finite $x$ keeps the stationary law broad. Because of the transient region near $t\sim T$, the boundary distribution of $\mathsf{r}(T)$ relevant for local observables differs from the bulk stationary law, see Fig.~\ref{fig:panel}b. Nonetheless, by isotropy, the full distribution of a spin component follows from the boundary law of $\mathsf{r}(T)$.
In Fig.~\ref{fig:mfexactdist}, we compare the distribution extracted in the $N\to\infty$ limit with the exact finite-$N$ dynamics (see also Appendix~D).

\paragraph{Purification dynamics. ---}
We now comment about the calculation of the purity. An exact computation of the free-energy barrier $\mathcal{I}^{*}$ connecting two vacua, the identity and the swap $(1,2)$ requires an Ansatz for $X^{\sigma\sigma'}$ more general than Eq.~\eqref{eq:replicaansatz}. An upper bound can be obtained by approximating the instanton as an abrupt jump between the two bulk sectors. This gives~\cite{SM}
\begin{equation} \label{main:purity}
e^{-\mathcal{I}^{*}}  \gtrsim 1 + (x^\ast)^2 -x^\ast k(x^\ast) (x^\ast k(x^\ast)+1).
\end{equation}
This approximation should improve when the instanton width is negligible compared with the long bulk interval, most clearly for $\gamma \to 0$, where $x^\ast \sim 3/(8\gamma) $ and $\mathcal{I}^\ast = \ln 2 - 4 \gamma + O(\gamma^{2})$ approaches its maximal value.
In the opposite limit $\gamma \gg 1$, $x^\star$ vanishes exponentially and $\mathcal{I}^{*} \leq \gamma e^{- 4 \gamma}$: even if it does not vanish, for large $\gamma$ the barrier becomes exponentially small. Such a rapid decay makes numerical verification of the absence of a phase transition arduous, since at sufficiently large $\gamma$, to see an exponentially increasing purification time would require reaching sizes $N \gg \mathcal{I}^\ast \sim e^{4 \gamma}$.

\paragraph{Conclusions. ---} We presented the exact solution of an interacting many-body spin system under continuous isotropic monitoring, finding that it does not undergo a MIPT: starting from the fully mixed state, the system always purifies in a time exponentially long in system size, i.e. the Landau-Ginzburg action is always in a replica-broken phase. This is a consequence of the analytic continuation to $n \rightarrow 1$: for larger $n$, the system does exhibit a disordered (permutation-symmetric) phase for sufficiently large $\gamma > \gamma_c(n)$.
A natural question is whether the logarithmic non-analyticity emerging at $n\to1$ is a general feature of monitored systems rather than a peculiarity of this solvable model; if so, it would signal a limitation of replica Landau-Ginzburg and perturbative RG approaches based on a regular expansion around the unbroken saddle~\cite{nahum2021measurement,nahum2023renormalization}, and it becomes important to understand whether it persists at finite spatial dimension. A further question is the role of the statistical rotational invariance: isotropy alone cannot explain the disappearance of the transition---the same model still displays transitions at integer $n>1$---but it may constrain the effective long-wavelength theory by enforcing additional degeneracies in the replica limit. Finally, an $N$-dependent scaling of $\gamma$ might lead to a non-conventional transition, as in many-body localisation~\cite{PhysRevB.102.125134} and integrability breaking~\cite{PhysRevB.105.214308}.

\paragraph{Acknowledgements.}
We are indebted to Denis Bernard, Pierre Le Doussal, Lorenzo Correale, Chris Baldwin, Jacopo De Nardis, David Huse and in particular Adam Nahum for useful discussions. We thank the Institut Pascal (University of Paris Saclay) and LPTMS for hospitality and support during the "Dynamical Foundations of Many-Body Quantum Chaos" and "OpenQMBP2023" programmes. The authors acknowledge support by the ANR JCJC grant ANR-21-CE47-0003 (TamEnt).

\let\oldaddcontentsline\addcontentsline
\renewcommand{\addcontentsline}[3]{}

\let\addcontentsline\oldaddcontentsline

\newpage
\clearpage
\appendix
\renewcommand{\thesection}{\Alph{section}}
\renewcommand{\thesubsection}{\thesection.\arabic{subsection}}
\setcounter{equation}{0}
\setcounter{figure}{0}
\setcounter{table}{0}
\renewcommand{\thetable}{\thesection\arabic{table}}
\renewcommand{\theequation}{\thesection.\arabic{equation}}
\renewcommand{\thefigure}{\thesection\arabic{figure}}
\renewcommand{\theHequation}{\thesection.\arabic{equation}}
\renewcommand{\theHfigure}{\thesection\arabic{figure}}
\makeatletter
\renewcommand{\p@section}{}
\renewcommand{\p@subsection}{}
\renewcommand{\p@equation}{}
\renewcommand{\p@figure}{}
\makeatother
\setcounter{secnumdepth}{2}

\begin{center}
{\Large End Matter}
\end{center}

\section{From Trajectory Dynamics to the Replica Formulation}
For continuous homodyne monitoring, the normalized conditional state $\rho_M(t)$
obeys a stochastic evolution driven by the measurement record
$M_i^\alpha(t)$. Because of Born rule,
the measurement record has a bias towards the quantum expectation value
\begin{equation}
\label{eq:signal}
    dM_i^\alpha = -2\sqrt{\gamma}\,\langle \hat S_i^\alpha\rangle dt + dY_i^\alpha,
\end{equation}
with $\langle \hat S_i^\alpha\rangle=\Tr[\rho_M \hat S_i^\alpha]$ and
$dY_i^\alpha dY_j^\beta = dt\,\delta_{ij}\delta^{\alpha\beta}$. Using this decomposition, one writes an evolution equation for $\rho_M$, including the Born-rule weights. It is known as the Stochastic Schr\"odinger equation and reads
\begin{equation}
\label{eq:SSE-end}
    d\rho_M
    =
    [d\rho]_{\rm uni}
    + \gamma dt\sum_{i,\alpha}\mathcal D_{\hat S_i^\alpha}[\rho_M]
    + \sqrt{\gamma}\sum_{i,\alpha} dY_i^\alpha
    \{\delta \hat S_i^\alpha,\rho_M\},
\end{equation}
where $[d\rho]_{\rm uni}$ contains the unitary part of the evolution \eqref{eq:rhouni}, $\delta \hat S_i^\alpha=\hat S_i^\alpha-\langle \hat S_i^\alpha\rangle$ and
$\mathcal D_{\hat O}[\rho]=\hat O\rho\hat O-\frac12\{\hat O^2,\rho\}$. The
discrete-time ancilla construction leading to these equations is recalled in the
Supplemental Material~\cite{SM}.

Eq.~\eqref{eq:SSE-end} is non-linear as expected. To obtain a linear dynamics, we introduce the unnormalized state
\begin{equation}
    \tilde\rho_M+d\tilde\rho_M
    =
    e^{-i d\hat H}\tilde\rho_M e^{i d\hat H^\dagger},
\end{equation}
with the non-Hermitian increment \eqref{eq:hamfull} playing the role of infinitesimal Kraus operator.
Consistently with Eq.~\eqref{eq:Frhoreplica}, the physical state is recovered as
\begin{equation}
    \rho_M(t)=\frac{\tilde\rho_M(t)}{\tr\tilde\rho_M(t)},   \quad  P[M]=\tr\tilde\rho_M(t)\,P_G[M] \;,
\end{equation}
with $P[M]$ the probability of the record $M$ and $P_G[M]$ is the unbiased Gaussian measure for the signals $dM_i^\alpha$.
\section{Replica Ansatz and Reduction to a Scalar Action}
This appendix justifies the symmetries of the replicated single-site action and the replica Ansatz they imply, and records the operator form of the saddle equations used in the main text. We work throughout with the left/right multiplication superoperators of the main text, $\hat S^\alpha_{a,+}[\rho]=\hat S^\alpha_a\rho$ and $\hat S^\alpha_{a,-}[\rho]=\rho\,\hat S^\alpha_a$, so that the generator $\mathcal L^{(n)}_X(t)$ of Eq.~\eqref{main:mathLn} acts directly on the replicated single-site operator $\trho1^{(n)}$; an equivalent vectorized (Liouville-space) formulation, useful for contact with part of the literature, is given in Sec. S2.2 of \cite{SM}. We denote by
\begin{equation}
\label{app:single-site-prop}
    \mathcal U^{(n)}_{X;t_2,t_1}
    =
    \mathcal T\exp\!\left[\int_{t_1}^{t_2}dt\,\mathcal L^{(n)}_X(t)\right]
\end{equation}
the corresponding time-ordered propagator, so that $\trho1^{(n)}(t_2)=\mathcal U^{(n)}_{X;t_2,t_1}[\trho1^{(n)}(t_1)]$. The large-\(N\) saddle equations identify the fields with single-site bilinears evaluated in the boundary-value problem fixed by the initial state $\rho_0=\mathbb 1/2$ and the boundary operator $\Omega^{(n)}$:
\begin{align}
\label{app:saddleX-general}
    X^{\sigma\sigma'}_{ab}(t)
    &=
    \frac{
    \Tr\bigl[\Omega^{(n)}\,\mathcal U^{(n)}_{X;T,t}\bigl[(\hat{\mathbf S}_{a,\sigma}\!\cdot\!\hat{\mathbf S}_{b,\sigma'})\,\mathcal U^{(n)}_{X;t,0}[\rho_0]\bigr]\bigr]}
    {\Tr\bigl[\Omega^{(n)}\,\mathcal U^{(n)}_{X;T,0}[\rho_0]\bigr]}\;,
\end{align}
where $\hat{\mathbf S}_{a,\sigma}\!\cdot\!\hat{\mathbf S}_{b,\sigma'}=\sum_\alpha \hat S^\alpha_{a,\sigma}\hat S^\alpha_{b,\sigma'}$ denotes the composition of the corresponding multiplication superoperators, inserted at the intermediate time $t$.
The symmetries of the original action \((S_n\times S_n)\rtimes\mathbb Z_2\) appear as transformations
connecting boundary-value problems in Eq.~\eqref{app:saddleX-general}. Whenever the boundary data are invariant
under the transformation, the transformed field is another solution of the same saddle equations; otherwise it gives the
corresponding solution in the transformed sector. The two permutation groups $P_{\sigma=\pm} \in S_n$ relabel
independently the replicas on the two contours,
\begin{equation}
\begin{aligned}
    &X^{\sigma\sigma'}_{ab}(t)
    \mapsto
    X^{\sigma\sigma'}_{P_\sigma(a)\,P_{\sigma'}(b)}(t)\;,\\
    &\rho_0\mapsto U_{P_+}\rho_0\,U_{P_-}^{-1}\;,
    \quad
    \Omega^{(n)}\mapsto U_{P_+}\Omega^{(n)}\,U_{P_-}^{-1}\;,
\end{aligned}
\end{equation}
where $U_P$ is the unitary that permutes the $n$ replicas according to $P$.
The \(\mathbb Z_2\) is the contour-exchange map, implemented by the \emph{antilinear} Hermitian-conjugation map \(\mathcal C[\rho]=\rho^\dagger\). Since the spin operators are Hermitian, it exchanges left and right multiplication,
\(\mathcal C\,\hat S^\alpha_{a,\sigma}\,\mathcal C=\hat S^\alpha_{a,-\sigma}\)
(e.g.\ \(\mathcal C\,\hat S^\alpha_{a,+}\,\mathcal C[\rho] = (\hat S^\alpha_a \rho^\dagger)^\dagger = \rho\,\hat S^\alpha_a = \hat S^\alpha_{a,-}[\rho]\)),
and acts as
\begin{equation}
\begin{aligned}
    &(\mathcal C X)^{\sigma\sigma'}_{ab}(t)
    =
    \bigl(X^{-\sigma,-\sigma'}_{ab}(t)\bigr)^* \;,\\
    &\rho_0\mapsto \rho_0^\dagger\;,\quad
    \Omega^{(n)}\mapsto \Omega^{(n)\dagger}\;.
\end{aligned}
\end{equation}
A solution that is symmetric under this $\mathbb{Z}_2$ group and real must satisfy $X^{++} = X^{--}$ and $X^{+-}_{ab} = X^{-+}_{ab} = X^{+-}_{ba}$. Admitting the possibility of a symmetry breaking as described in Eq.~\eqref{eq:symmbreakP} leads to the Ansatz \eqref{eq:replicaansatz}.
\subsection{Reduction to a tridiagonal matrix
\label{app:redtrid}
}
After inserting \eqref{eq:replicaansatz} into the generator $\mathcal L^{(n)}_X$ of Eq.~\eqref{main:mathLn},
the static bulk generator becomes
\begin{equation}
\label{app:mathLnqr-end}
\begin{split}
    \mathcal L_X^{(n)}(\trho1^{(n)})
    =
    (\gamma+r) \hat S^{\alpha}_{\rm tot}\trho1^{(n)} \hat S^{\alpha}_{\rm tot}
    + \frac{1}{2}(\gamma-q)
    \left\{\hat{\mathbf S}^2_{\rm tot},\trho1^{(n)}\right\} \\
    \qquad
    + X\sum_a \hat S^\alpha_a\trho1^{(n)} \hat S^\alpha_a
    + \frac{3n}{4}\left(q-2\gamma-\frac34\right)\trho1^{(n)},
\end{split}
\end{equation}
where \(\mathbf S_{\rm tot}=\sum_a\mathbf S_a\).
The sectors are labeled by $\ell=(n\!\!\mod 2)/2,\ldots,n/2$, and we normalize the projectors in the Hilbert-Schmidt norm, $\Tr^{(n)}[\Pi_\ell\Pi_{\ell'}]=\delta_{\ell,\ell'}$.
As explained in the main text,
rotational symmetry and the choice of the initial state reduce the diagonalization to sectors of fixed total spin
\(\mathbf S_{\rm tot}^2=\ell(\ell+1)\): indeed the infinite-temperature initial state expands on this basis as $\tilde\varrho^{(n)}(0)=\mathbb{1}/2=\frac12\sum_\ell\Tr^{(n)}[\hat\Pi_\ell]\,\hat\Pi_\ell$.
Projecting the generator on
this orthonormal operator basis gives Eq.~\eqref{eq:Lprojll}, where
\begin{align}
\label{app:Ldef-end}
    L_{\ell,\ell'}
    &=
    \ell(\ell+1)\delta_{\ell,\ell'}
    + x\sum_{a,\alpha}\Tr\!\left[
    \Pi_\ell \hat S_a^\alpha \Pi_{\ell'} \hat S_a^\alpha
    \right]\;,
\end{align}
with $x=X/(2\gamma+r-q)$. The largest eigenvalue of $(\mathcal{L}_X^{(n)})_{\ell, \ell'}$ is
\[
    \lambda_n
    =
    (2\gamma+r-q)\Lambda_n(x)
    +\frac{3n}{4}\left(q-2\gamma-\frac34\right),
\]
where \(\Lambda_n(x)\) is the largest eigenvalue of \(L\).  For spin \(1/2\),
\(L\) is tridiagonal, with nonzero elements
\begin{align}
\label{eq:Lmatrdef}
    L_{\ell,\ell}
    &=\ell(\ell+1)+\frac12\left(\frac n2+1\right)x,
    \qquad (\ell\neq0), \nonumber\\
    L_{\ell,\ell+1}&=L_{\ell+1,\ell}
    =\frac{x}{2}
    \sqrt{\left(\frac n2-\ell\right)\left(\frac n2+\ell+2\right)},
\end{align}
and \(L_{00}=0\) for even \(n\) (see~\cite{SM} for the derivation).

\begin{figure}[t]
    \includegraphics[width=0.75\columnwidth]{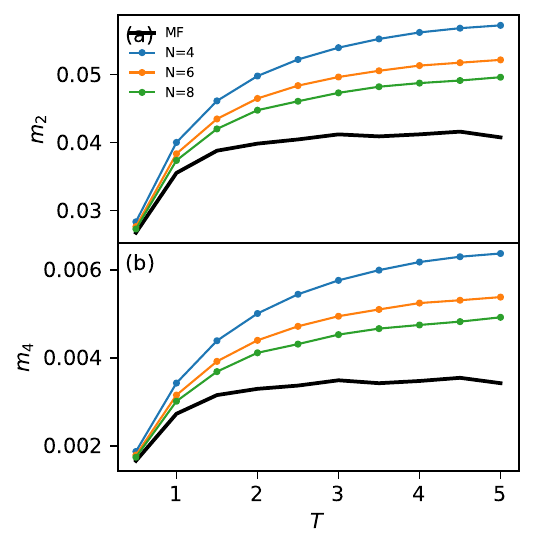}
    \par\smallskip
    \caption{Time dependence of $m_{2,4}(T)$ predicted by the self-consistent single-spin dynamics and from exact trajectory simulations of the full monitored density matrix at $\gamma=0.38$.}
    \label{fig:mfexactmom}
\end{figure}
Finally, in the limit $T\to\infty$, with the Ansatz \eqref{eq:replicaansatz}, the action \eqref{eq:Idefmain} acquires the static form
\begin{equation}
    \mathcal I(q,r,X)
    =
    \frac{n(n-1)}{2}(r^2-q^2)
    +\frac{n}{2}(r+X)^2
    -\frac{9n}{32}
    -\lambda_n,
\end{equation}
where the constant $-9n/32$ originates from the diagonal entries $\bar X^{++}_{aa} = 3/4$ in the quadratic part of Eq.~\eqref{eq:Idefmain} and ensures $\mathcal I \equiv 0$ at $n=1$.
Its stationarity conditions with respect to $q,r,X$ are
\begin{align}
    n(n-1)q-\Lambda_n(x)+x\Lambda_n'(x)+\frac{3}{4}n&=0, \nonumber\\
    n(nr+X)-\Lambda_n(x)+x\Lambda_n'(x)&=0, \label{app:saddleqrX}\\
    n(r+X)-\Lambda_n'(x)&=0. \nonumber
\end{align}
These conditions can be used to fix $q$ and $r$ and, through the last one, re-express the spectral derivative $\Lambda_n'(x)$ in terms of $\Lambda_n(x)$ itself. In this way, the action collapses to a function of the single reduced variable $x$, entering only through $\Lambda_n(x)$,
$\mathcal I^{\rm full}_{n}(x) = \mathcal I_{n}(x) - \gamma n(n-1)(4 \gamma n + 3)/2$, with $\mathcal{I}_n(x)$ given in Eq.~\eqref{main:mathIx}~(see also S3.1 in \cite{SM}).

\section{Comparison with Exact Dynamics}
The single-spin dynamics provides a direct numerical prediction for local observables at finite monitoring time $T$. Denoting as $\mathsf{r}_\alpha$ the components of the Bloch vector $\mathsf{r}(T)$ and by  $P_T(\mathsf{r})$ the distribution of its modulus $\mathsf{r}(T)$, we have $\langle \hat S^\alpha \rangle := \tr[\varrho(T) \hat S^\alpha] = \mathsf{r}_\alpha/2$ and
\begin{equation}
    \begin{aligned}
    m_k(T) := \mathbb{E}[\langle S_\alpha \rangle^k]
    = \frac{E[\mathsf{r}^k(T)]}{2^{k}(k+1)}
     \;.
    \end{aligned}
\end{equation}
where $E[\ldots]$ denotes averaging wrt $P_T(\mathsf{r})$. Here, we used that, because of isotropy, the components $r_\alpha$ are uniformly distributed on $[-\mathsf{r}, \mathsf{r}]$ once conditioned on the modulus $\mathsf{r}$. In particular, \(m_2(T)=E[\mathsf{r}^2(T)]/12\) and \(m_4(T)=E[\mathsf{r}^4(T)]/80\).
More generally, we approximate the full distribution of $\langle S_\alpha \rangle$ from the sampled moduli as
\begin{equation}
    \begin{split}
    P_T(\langle S^\alpha\rangle)
    &\simeq
    \frac{1}{N_{\rm samp}}
    \sum_{j=1}^{N_{\rm samp}}
    \frac{1}{\mathsf{r}_j}\,
    \Theta\!\left(
    \frac{\mathsf{r}_j}{2}-|\langle S^\alpha\rangle|
    \right).
    \end{split}
\end{equation}
These predictions can be compared directly with exact trajectory simulations of the full density matrix, averaged over sites, spin components, and samples. The full distribution of a spin component is shown in the main text, Fig.~\ref{fig:mfexactdist}. In Fig.~\ref{fig:mfexactmom}, we report the time dependence of $m_{2,4}(T)$. Although finite-size corrections are still present, the finite-$N$ curves approach the mean-field prediction monotonically.

\clearpage
\onecolumngrid
\setcounter{page}{1}
\setcounter{section}{0}
\setcounter{subsection}{0}
\setcounter{equation}{0}
\setcounter{figure}{0}
\setcounter{table}{0}
\renewcommand{\thesection}{S\arabic{section}}
\renewcommand{\thesubsection}{\thesection.\arabic{subsection}}
\renewcommand{\thetable}{S\arabic{table}}
\renewcommand{\theequation}{\thesection.\arabic{equation}}
\renewcommand{\thefigure}{S\arabic{figure}}
\renewcommand{\theHequation}{\thesection.\arabic{equation}}
\renewcommand{\theHfigure}{S\arabic{figure}}
\makeatletter
\renewcommand{\p@section}{}
\renewcommand{\p@subsection}{}
\renewcommand{\p@equation}{}
\renewcommand{\p@figure}{}
\makeatother

\begin{center}
{\Large Supplementary Material \\
\titleinfo
}
\end{center}
\section{Derivation of the weak measurement equations \label{sec:derweak}}
\subsection{Repeated interaction with the ancilla}
\noindent
Here, we describe briefly how the stochastic Schr\"odinger equation (SSE) for the weak measurement dynamics can be derived.
To simplify the notation, we will consider a generic system evolving under the (Hermitian) Hamiltonian $\hat H_0$ and where a single observable $\hat O$ is continuously monitored; the symbol $\hat H$ is reserved for the full non-Hermitian generator including the measurement terms, see Eq.~\eqref{eq:timeevol} below. First of all, we discretize time considering a finite small interval $\Delta t$. 
The basic idea is that in a time step $t \in [\tind \Delta t, (\tind+1) \Delta t]$, $\tau \in \mathbb{N}$, the system $\mathcal{S}$ is coupled to a (new) ancilla $\mathcal{A}$. For simplicity, the ancilla is supposed to be a spin $1/2$ initial set in the state
\begin{equation}
    \ket{\mathcal{A}} = \frac{\ket{+} + \ket{-}}{\sqrt{2}}
\end{equation}
At time $t = \tind \Delta t$, the state of the system + ancilla is thus in the factorized state
\begin{equation}
    \ket{\Psi_{\tind}} = \ket{\psi_\tind} \otimes \ket{\mathcal{A}}
\end{equation}
The time evolution up to the next time step $\tind+1$ is performed in two steps:
\begin{enumerate}
    \item the system and the ancilla evolve unitarily for a time $\Delta t$ and get entangled because of the coupling between them;
    \item the $z$-component of the ancilla spin is measured projectively; 
\end{enumerate}    
Let's analyse the two steps. Under the unitary evolution $\hat U$, one arrives at
\begin{equation}
\label{eq:decompPsi}
    \ket{\Psi_{\tind}}' = \hat U \ket{\Psi_{\tind}} = (\hat K_+ \ket{\psi_\tind}) \otimes \ket{+} +
    (\hat K_- \ket{\psi_\tind}) \otimes \ket{-} \;, \qquad \hat K_{\pm} = \braket{\pm | \hat U | \mathcal{A}}
\end{equation}
where we introduced the operators $\hat K_{\pm}$ acting on the Hilbert space of the system $\mathcal{S}$. Unitarity of $\hat U$ and normalization of $\ket{\mathcal{A}}$ imply the constraint
\begin{equation}
    \ide_\mathcal{S} = \braket{A | \hat U^\dag \hat U | A}  = \hat K_+^\dag \hat K_+ + \hat K_-^\dag \hat K_-
\end{equation}
where $\ide_{\mathcal{S}}$ is the identity operator on the system Hilbert space. This shows that $\hat K_{\pm}$ are Kraus operators.

Measuring the spin $\hat \sigma_z$ of the ancilla, one can obtain two possible outcomes $a_\tind = \pm 1$.
Correspondingly, the state of the system takes the form
\begin{equation}
\label{eq:projK}
    \ket{\psi_{\tau+1}} = \frac{\hat K_{a_\tind} \ket{\psi_\tind}}{\sqrt{\braket{\psi_\tind | \hat K_{a_\tind}^\dagger \hat K_{a_\tind} \ket{\psi_\tind}}}}
\end{equation}
\subsection{Continuous time limit}
\noindent
For the continuous limit $\Delta t \to 0$, we need an explicit form of the operators $\hat K_{\pm}$. For simplicity, we focus on the situation when only one operator $\hat O$ is being monitored, the generalisation being straightforward.
Let us first consider the explicit form of the system and ancilla Hamiltonian $\mathcal{S} + \mathcal{A}$. We take
\begin{equation}
    \hat U = e^{- \imath \Delta t \hat H_{\mathcal{S} + \mathcal{A}}} \;, \quad
    \hat H_{\mathcal{S} + \mathcal{A}} = \hat H_0 + \lambda \hat O \hat \sigma_y \;.
\end{equation}
Choosing $\hat \sigma_y$ as the operator on the ancilla is the simplest way to achieve monitoring.
The scaling limit is achieved taking $\lambda \to \infty$ and $\Delta t \to 0$ but in such a way that $\gamma = \lambda^2 \Delta t$ is kept constant where the rate $\gamma$ parameterises the strength of the measurements. With this choice, expanding to the order $O(\Delta t)$, we have
\begin{equation}
    \hat U = e^{-\imath (\Delta t \hat H_0 + \sqrt{\gamma \Delta t} \hat O \hat \sigma_y)} \sim \ide - \imath \Delta t \hat H_0 - \imath \sqrt{\gamma \Delta t} \hat O \hat \sigma_y - \frac{1}{2} \gamma \Delta t \hat O^2  + O (\Delta t^{3/2})
\end{equation}
From the definitions of $\hat K_{\pm}$, we thus obtain
\begin{equation}
    \hat K_{\pm} = \frac{1}{\sqrt{2}} \left(\ide - \imath \Delta t \hat H_0 \mp \sqrt{\gamma \Delta t} \hat O - \frac1 2 \gamma \Delta t \hat O^2  \right) + O(\Delta t ^{3/2})
\end{equation}
In order to compute the norm, we expand
\begin{equation}
    \hat K_{a}^\dagger \hat K_{a} = \frac 12 - a \sqrt{\gamma \Delta t} \hat O + O(\Delta t^{3/2})
\end{equation}
Therefore, we have
\begin{multline}
    \ket{\psi_{\tind+1}} = \ket{\psi_{\tind}} - \imath \hat H_0 \Delta t \ket{\psi_{\tind}} - a_\tind \sqrt{\gamma \Delta t} ( \hat O - \langle \hat O \rangle) \ket{\psi_\tind} + \frac 32  \gamma \Delta t \langle \hat O\rangle^2 \ket{\psi_{\tind}}  - \gamma \Delta t \langle \hat O  \rangle \hat O \ket{\psi_{\tind}} - \frac 1 2 \gamma \Delta t \hat O^2 \ket{\psi_{\tind}} + O(\Delta t^{3/2})
\end{multline}
where we have introduced the notation $\langle \hat O \rangle = \braket{\psi_\tind | \hat O | \psi_\tind}$.
Finally, to get the continuous time limit as a stochastic differential equation, we observe that the measurement outcome $a_\tau$ is a random variable that satisfies
\begin{equation}
    \overline{a_\tind} = - 2 \sqrt{\gamma \Delta t} \langle \hat O \rangle  \;, \quad \overline{a_\tind^2} = 1 \;.
\end{equation}
Therefore, defining $M_\tind =  \sqrt{\Delta t} \sum_{\tind'\leq\tind} a_{\tind'}$, we find that the variable $M_\tau$ converges in the limit $\Delta t \to 0$ to a stochastic process solving
\begin{equation}
\label{eq:dY}
    dM = - 2 \sqrt{\gamma} \langle \hat O \rangle dt + dY
\end{equation}
being $Y_t$ a standard Wiener process (i.e. $\overline{dY} = 0$ and $\overline{dY^2} = dt$). 
Eq.~\eqref{eq:dY} is easily generalized to the multi-spin signal equation~\eqref{eq:signal}.
Deriving the SSE can be achieved by formally replacing $a_\tind \to dM/\sqrt{\Delta t}$ and using \eqref{eq:dY} in the limit $\Delta t \to 0$, we recover
\begin{equation}
\label{eq:SSE-supp}
 d\ket{\Psi_t} = -\imath \hat H_0 dt \ket{\Psi_t} + \left(\sqrt{\gamma} [ \hat{O} - \langle \hat O \rangle_t] dY - \frac \gamma 2 [ \hat O - \langle \hat O \rangle_t]^2 dt \right) \ket{\Psi_t} \;.
\end{equation}
This equation can be easily generalised to the case where one simultaneously monitors several observables $O_1,\ldots$ 
introducing independent Wiener processes for each operator undergoing monitoring
\begin{equation}
\label{eq:SSEmany}
 d\ket{\Psi_t} = -\imath \hat H_0 dt \ket{\Psi_t} + \sum_i \left(\sqrt{\gamma} [ \hat{O}_i - \langle \hat O_i \rangle_t] dY_i - \frac \gamma 2 [ \hat O_i - \langle \hat O_i \rangle_t]^2 dt \right) \ket{\Psi_t}
\end{equation}
From this, one can derive the evolution for the density matrix $\rho = \ket{\Psi_t}\bra{\Psi_t}$
\begin{equation}
\label{eq:lindblbiasmeas}
d\rho = -\imath dt [\hat H_0, \rho] -
\gamma\sum_i \left[\frac 1 2 \{\hat O_i^2, \rho\} - \hat O_i \rho \hat O_i\right]
+ \sqrt{\gamma}\sum_i dY_i \{\hat O_i - \langle \hat O_i \rangle_t, \rho\}
\end{equation}
Generally speaking, we will refer to these two equations as the stochastic Schrödinger equations (SSE).

\subsection{Replica trick \label{sec:replicatrick}}
The previous derivation allows us to compute averages over quantum trajectories of the density matrix. In the main text, we considered the situation where even the unitary part of the dynamics induced by the Hamiltonian is stochastic. Formally that amounts to replace $dt \hat {H}_0$ with $d\hat{H}_0$ the Hamiltonian increment, and including the $O(d\hat{H}_0^2)$ term in agreement with Ito's calculus (see Eq.~\eqref{eq:rhouni} in the main text). Here, for simplicity of the notation, we only consider averaging over the measurement outcomes assuming only one operator is being monitored. The most general case is simply achieved by redefining the average over trajectories.

Typical examples of quantities of interest are moments of observables evaluated on the normalized state at the final monitoring time $T$:
\begin{equation}
\label{eq:momave}
    \mathbb{E}\!\left[\tr(\hat A\rho(T))^k\right]
    =
    \int \mathcal{D} Y P_{W}(Y)\,
    \tr(\hat A\rho_{Y}(T))^k
\end{equation}
where $\mathbb{E}[\ldots]$ indicates an averaging over the measurement outcomes,
and $P_W(Y)$ is the measure over the Wiener process introduced in Eq.~\eqref{eq:dY}. It is also useful to rewrite Eq.~\eqref{eq:momave} in a different way. Going back to the problem in discretized time, let $\mathsf{T}=T/\Delta t$ be the number of measurement time steps. In agreement with Eq.~\eqref{eq:projK}, we can write the unnormalized density matrix associated with a measurement record $\mathbf a=(a_1,\ldots,a_{\mathsf{T}})$ as
\begin{equation}
    \tilde\rho_{\mathbf a}
    =
    \mathsf{K}_{\mathbf a}\rho_0\mathsf{K}_{\mathbf a}^{\dagger}
    \;,\qquad
    \rho_{\mathbf a}
    =
    \frac{\tilde\rho_{\mathbf a}}{\tr\tilde\rho_{\mathbf a}}
    \;,\qquad
    \mathsf{K}_{\mathbf a} = \hat K_{a_{\mathsf{T}}} \ldots \hat K_{a_2} \hat K_{a_1} \;,
\end{equation}
where $\mathsf{K}_{\mathbf{a}}$ denotes the Kraus operator associated to the full sequence of measurement outcomes (the dependence on the final time $T$ is left implicit).
The $a_\tind$'s indicate the outcomes of the measurements on each ancilla. The Born-rule probability of the record is $P(\mathbf a)=\tr[\tilde\rho_{\mathbf a}]$, while the normalized state is $\rho_{\mathbf a}$. Then, we have
\begin{equation}
\label{eq:obsfull}
    \mathbb{E}\!\left[\tr(\hat A\rho(T))^k\right]
    =
    \sum_{\mathbf a} P(\mathbf a)
    \left(
    \frac{\tr(\hat A\tilde\rho_{\mathbf a})}{\tr\tilde\rho_{\mathbf a}}
    \right)^k
    =
    \sum_{\mathbf a}
    \tr(\tilde\rho_{\mathbf a})^{1-k}
    \tr(\hat A\tilde\rho_{\mathbf a})^k \;.
\end{equation}
This form makes the Born-rule reweighting explicit. Additionally, we can rewrite in the limit of small $\Delta t$
\begin{equation}
\label{eq:expB}
    \hat K_a = \frac{1}{\sqrt{2}} \exp[ -\text{i}\Delta t \hat H_0 - a \sqrt{\gamma \Delta t} \hat O - \gamma \Delta t \hat O^2] + O(\Delta t^{3/2})
\end{equation}
where, because of Ito's calculus, the last term in the exponent must be inserted to obtain the proper expansion up to the order $O(\Delta t)$. We can now take the limit $\Delta t \to 0$. As before, we set $M_\tind =  \sqrt{\Delta t} \sum_{\tind'\leq\tind} a_{\tind'}$ and are interested in considering the continuous time limit $\Delta t \to 0$. Since the Born-rule probability is now carried explicitly by the factors of $\tr\tilde\rho_{\mathbf a}$ in Eq.~\eqref{eq:obsfull}, we can rewrite the $2^{-\mathsf{T}} \sum_{\mathbf a} (\ldots)$ as an unbiased average over the measurement outcomes. The factor $2^{\mathsf{T}}$ can be absorbed in the Kraus operators. Explicitly, in the limit $\Delta t \to 0$,
\begin{equation}
 2^{-\mathsf{T}} \sum_{\mathbf{a}} (\ldots) = \int \mathcal{D}M P_W(M) (\ldots) \;, \quad \mathcal{K}_{T, M}\equiv\lim_{\Delta t \to 0} 2^{\mathsf{T}/2} \mathsf{K}_\textbf{a}
\end{equation}
where $P_W(M)$ is the measure of a standard Wiener process. Consistently with the main text (see Eq.~\eqref{eq:Frhoreplica}), we denote as $\mathbb{E}_G[\ldots]$ the corresponding expectation values, with  $\mathbb{E}_G[dM] = 0$ and $\mathbb{E}_G[dM^2] = dt$.  Thus, we can write
\begin{equation}
\label{eq:obsfulllim}
    \mathbb{E}\!\left[\tr(\hat A\rho(T))^k\right]
    =
    \int \mathcal{D}M P_{W}(M)
    \tr(\tilde\rho_{M})^{1-k}
    \tr(\hat A\tilde\rho_{M})^k =: \mathbb{E}_G(\tr(\tilde\rho_{M})^{1-k}
    \tr(\hat A\tilde\rho_{M})^k) \;,
    \
    \tilde\rho_{M}=\mathcal{K}_{T,M}\rho_0\mathcal{K}_{T,M}^\dagger .
\end{equation}
Note that in the continuous limit the non-unitary evolution operator $\mathcal{K}_{t, M}$, with $t \in [0,T]$ an intermediate time, satisfies
\begin{equation}
\label{eq:timeevol}
    \mathcal{K}_{t + dt, M}= 
    e^{- \imath d \hat{H}} \mathcal{K}_{t, M} \;, \qquad 
    d \hat{H} =  \hat{H}_0 dt - \imath \sqrt{\gamma} dM \hat{O} - \imath \gamma  \hat{O}^2 dt
\end{equation}
so that $\hat{H}$ plays the role of a non-hermitian stochastic Hamiltonian. Its generalisation to many monitored operators and a stochastic unitary dynamics (note that in this case the Ito's term shown in Eq.~\eqref{eq:rhouni} shall not be included, as we are writing the evolution in exponential form) leads to Eq.~\eqref{eq:hamfull} in the main text.

Finally, in order to avoid the presence of a negative exponent $(1-k)$ in the averaging procedure in Eq.~\eqref{eq:obsfulllim}, 
we can make use of the replica trick, replacing $1-k$ with $n-k$. For integer $n$, we define the averaged density matrix for $n$ replicas as
\begin{equation}
    \rho^{(n)}
    =
    \int \mathcal{D} M P_{W}(M)\,
    \underbrace{\tilde\rho_{M} \otimes \ldots \otimes \tilde\rho_{M} }_{n}
    \equiv
    \mathbb{E}_G[\tilde\rho_{M}^{\otimes n}] \;.
\end{equation}
In this way, we can write via the replica trick
\begin{equation}
    \mathbb{E}\!\left[\tr(\hat A\rho(T))^k\right]
    =
    \lim_{n \to 1}
    \Tr^{(n)}\!\left[
    \rho^{(n)}
    \left(\hat A^{\otimes k}\otimes \ide^{\otimes(n-k)}\right)
    \right] \;.
\end{equation}
More general observables, for instance those related to the spectrum of $\rho$, can be accessed by replacing $\hat A^{\otimes k}$ with an appropriate $\Omega_k$, as explained around Eq.~\eqref{eq:Frhoreplica} of the main text.

\section{Replica path integral}

\subsection{Derivation of the mean-field action}
\noindent
We consider the noisy non-Hermitian evolution
generated by $d\hat{H}$ in Eq.~\eqref{eq:hamfull} of the main text. In this subsection we derive Eqs.~\eqref{eq:Idefmain} and \eqref{main:mathLn}. In this section and in the following, we consider a generic value $s$ of the spin for each individual degree of freedom, as it clarifies the physical origin of some numerical coefficients. Eventually, we will set $s = 1/2$.
We want now to derive a path-integral representation for the expectation of a functional of the density matrix. As explained in the main text, using the replica trick it amounts to compute
\begin{equation}
\mathbb{E}[F[\rho]] = \lim_{n \to 1} \mathbb{E}_G[\Tr^{(n)}[\Omega^{(n)} \tilde\rho_M(T)^{\otimes n}]]
\end{equation}
for an appropriate choice of $\Omega^{(n)}$ which depends on the functional $F[\rho]$, see the case of the purity around Eq.\eqref{eq:gaussE} in the main text. To do so, we exploit the coherent-state path-integral approach. Given a unit vector
$\mathbf{n}(\theta, \phi) = (\sin \theta \cos \phi, \sin \theta \sin \phi, \cos \theta)$, we introduce the spin-coherent state of a single spin $s$ as the eigenstate in the positive direction of $\mathbf{n} \cdot \hat{\mathbf{S}}$ with $\hat{S} = (S^x, S^y, S^z)$. Explicitly, setting $\mathbf{s}(\theta, \phi) = s \mathbf{n}(\theta, \phi)$, we have
\begin{equation}
    \ket{\mathbf{s} (\theta, \phi)}  = \exp\left[-i\frac{\theta}{\sin\theta}(\vec{z} \times \mathbf{n}) \cdot  \hat{\mathbf{S}}\right] \ket{\uparrow}
\end{equation}
which leads to the resolution of the identity
\begin{equation}
    \frac{2s+1}{4\pi}\int d \omega \ket{\mathbf{s} (\theta, \phi)} \bra{\mathbf{s} (\theta, \phi)} = \ide
\end{equation}
where $d \omega = \sin\theta d\theta d\phi$ is the measure on the 2-sphere. We also have 
\begin{equation}
    \braket{\hat S^{\alpha}} \equiv \bra{\mathbf{s}} \hat S^{\alpha} \ket{\mathbf{s}} = s^{\alpha}
\end{equation}
We write the time evolution of the unnormalized density matrix as
\begin{equation}
    \tilde\rho(T) = e^{-\imath d\hat H} \ldots e^{-\imath d\hat H} \rho_0
    e^{\imath d\hat H^\dag } \ldots e^{\imath d\hat H^\dag } \;.
\end{equation}
Then, we consider the replicated many-body density matrix $\tilde \rho(T)^{\otimes n}$
and introduce a resolution of the identity between each time step, for each spin $i$ and replica $a$. 
We also use the additional $\pm$ notation to denote the forward/backward time evolution.
To avoid some subtleties of Ito's calculus in the path integral, which are the consequence of the non-smoothness of the Wiener processes $M_j^\alpha(t)$ and $W_{ij}^{\alpha \beta}(t)$, we consider a regularization procedure. We momentarily assume that the processes $M_j^\alpha(t)$ and $W_{ij}^{\alpha \beta}(t)$ are smooth and differentiable in time, setting  $dM_j^{\alpha} = m^{\alpha}_j (t) dt$,  $dW_{ij}^{\alpha, \beta} = w^{\alpha, \beta}_{ij} (t) dt$. In other words, we have the correlators
\begin{equation}
    \mathbb{E}_G[w_{ij}^{\alpha \beta}(t) w_{ij}^{\alpha \beta}(t')] = \delta_{\epsilon}(t-t') \;, \qquad
    \mathbb{E}_G[m_{i}^{\alpha}(t) m_{i}^{\alpha}(t')] = \delta_{\epsilon}(t-t') 
\end{equation}
where $\delta_\epsilon(t)$ is a smooth mollifier of the Dirac delta while the correlators vanish when any indices are different.
 Expanding for a short time interval $dt$
\begin{equation}
\begin{split}
    \bra{\mathbf{s} (t+dt)} &e^{- d \hat{H} }\ket{\mathbf{s}(t)} = \bra{\mathbf{s} (t+dt)} \exp{ \left[ - \text{i} \frac{J}{\sqrt{N}} \sum_{a,i<j} w^{\alpha \beta}_{ij} (t) S^{\alpha}_{i,a} S^{\beta}_{j,a} dt - \sqrt{\gamma} \sum_{a,j} m^{\alpha}_j (t) S^{\alpha}_{j,a} dt  - \gamma \sum_{a,j}  \mathbf{S}_{a,j}^2 dt \right] } \ket{\mathbf{s} (t)} \\ 
    &= 1 +  \sum_a \braket{\delta s | s_{j,a}} dt - \text{i} \frac{J}{\sqrt{N}} \sum_{a,i<j} w^{\alpha \beta}_{ij} (t) \braket{S^{\alpha}_{a,i} S^{\beta}_{a,j}} dt - \sqrt{\gamma} \sum_{a,j} m^{\alpha}_j (t) \braket{S^{\alpha}_{a,j}} dt - \gamma n N s(s+1) dt \\
    &= \text{exp} \left[ \sum_a \braket{\delta s | s_{j,a}} dt - \text{i} \frac{J}{\sqrt{N}}\sum_{a,i<j} w^{\alpha \beta}_{ij} (t) s^{\alpha}_{a,i} s^{\beta}_{a,j} dt - \sqrt{\gamma} \sum_{a,j} m^{\alpha}_j (t) s^{\alpha}_{a,j} dt - \gamma n N s(s+1) dt \right] , 
\end{split}
\end{equation}
where in the last line we implicitly assumed $i \neq j$, as the $i = j$ terms will only result in $O(N^{-1})$ corrections. In the continuum limit we get
\begin{equation}
\label{eq:pathintegralO}
     \Tr^{(n)}[\Omega^{(n)} \tilde\rho(T)^{\otimes n}] = \int \mathcal{D}[\mathbf{s}] \  e^{\mathcal{S}[\mathbf{s}]} \ \bra{\mathbf{s}^{-}_{a,j} (T)} \Omega^{(n)} \ket{\mathbf{s}^{+}_{a,j} (T)} \bra{\mathbf{s}^{+}_{a,j} (0)} \rho_0 \ket{\mathbf{s}^{-}_{a,j} (0)},
\end{equation}
where $\sigma = \pm$ distinguishes Keldysh contours and $\mathcal{D}[\mathbf{s}] = \prod_{a,j,\sigma} \prod_t d \omega_{a,j}^{\sigma} (t)$ denotes the product over time slices of the spherical measures $d\omega = \sin\theta \, d\theta \, d\phi$ associated with each spin-coherent vector $\mathbf{s}_{j,a,\sigma}(t)$, 
\begin{equation} 
\begin{split}
    \mathcal{S} = \sum^{n}_{a=1} &\sum_{\sigma = \pm} \left[ \text{i} \sigma \sum_j \mathcal{S}_{\rm top} [\mathbf{s}_{j,a \sigma}] - \text{i} \frac{J}{\sqrt{N}}\sigma \sum_{i<j} \int dt w_{i,j}^{\alpha \beta}(t) s^{\alpha}_{i,a \sigma} (t)  s^{\beta}_{j,a \sigma} (t)  -\sqrt{\gamma} \sum_j \int m_{j}^{\alpha}(t) s^{\alpha}_{j,a \sigma} (t)   \right] \\ &- 2 \gamma n N T s(s+1) , 
\end{split}
\end{equation}
and the topological term has the usual form derived from the coherent state overlap
\begin{equation}
    \mathcal{S}_{\rm top} [\mathbf{s}] := s \int dt \dot{\phi} \left( 1 - \cos \theta \right)
\end{equation}
Here $\Omega^{(n)}$ denotes the replica-space boundary insertion associated with the polynomial functional $F[\rho]$. Now we can perform the average over the realizations of both the noises. Taking the $\epsilon \rightarrow 0$ limit we get:
\begin{equation}   \label{EO2}
     \mathbb{E}_G[\Tr^{(n)}[\Omega^{(n)} \tilde\rho(T)^{\otimes n}]] = \int \mathcal{D}[\mathbf{s}] \  e^{\Bar{\mathcal{S}}[\mathbf{s}]} \ \bra{\mathbf{s}^{-}_{a,j} (T)} \Omega^{(n)} \ket{\mathbf{s}^{+}_{a,j} (T)} \bra{\mathbf{s}^{+}_{a,j} (0)} \rho_0 \ket{\mathbf{s}^{-}_{a,j} (0)},
\end{equation}
where we introduced the bilinears
\[
    B^{\sigma\sigma'}_{ab}(t):=
    \sum_j \mathbf s_{j,a\sigma}(t)\!\cdot\!\mathbf s_{j,b\sigma'}(t),
\]
so that
\begin{equation}
    \Bar{\mathcal{S}} [\mathbf{s}] =
    \text{i} \sum_j \sum_{a, \sigma} \sigma \mathcal{S}_{\rm top}[\mathbf{s}_{a,j,\sigma}]
    +  \frac{\gamma}{2} \sum_{ab,\sigma \sigma'} \int dt\, B^{\sigma\sigma'}_{ab}(t)
    - \frac{J^2}{4N}  \sum_{ab,\sigma \sigma'} \int dt\, \sigma \sigma^{\prime}
    \bigl(B^{\sigma\sigma'}_{ab}(t)\bigr)^2 . 
\end{equation}


After the noise average, the action is quadratic in the bilinears $B$, with an overall $1/N$ factor due to the all-to-all form of the interaction. It is then convenient to apply the Hubbard-Stratonovich identities to each independent matrix element of $B$,
\begin{equation}
        e^{A^2/4N} \sim \int d\mathcal{X}\, e^{-N \mathcal{X}^2/4 + \mathcal{X} A/2} \hspace{1cm} e^{-A^2/4N} = \int d\mathcal{X}\, e^{-N\mathcal{X}^2/4 + i\mathcal{X} A/2}
\end{equation}
using the first form for the mixed-contour channels and the second one for the same-contour channels. We denote by $\mathcal X^{\sigma\sigma'}_{ab}$ the resulting Hubbard-Stratonovich fields. In this way the term quadratic in $B$ is replaced by a quadratic form in $\mathcal X$ plus a linear coupling $\mathcal X\cdot B$. This gives
\begin{multline}
\label{eq:EOpathint}
    \mathbb{E}_G[\Tr^{(n)}[\Omega^{(n)} \tilde\rho(T)^{\otimes n}]] \sim \int \mathcal{D}[\mathcal X] e^{- \frac{N}{4} \int dt \sum_{a,b, \sigma, \sigma'} (\mathcal X^{\sigma \sigma'}_{a,b})^2}  \left( \int \mathcal{D}[\mathbf{s}] e^{\mathcal{S}^{(1)}[\mathbf{s}; \mathcal{X}]} \ \bra{\mathbf{s}^{\sigma}_{a,j} (T)} \Omega^{(n)} \ket{\mathbf{s}^{\sigma}_{a,j} (T)} \bra{\mathbf{s}^{+}_{a,j} (0)} \trho1^{(n)}(0) \ket{\mathbf{s}^{-}_{a,j} (0)} \right)^N \\ := \int \mathcal{D}[\mathcal X] e^{-NT\mathcal I[\mathcal X]}\; .
\end{multline}
where in the second line we implicitly defined the unreduced effective action $\mathcal I[\mathcal X]$ for the Hubbard-Stratonovich field $\mathcal X$. For clarity, we stress that the path-integral measure is over symmetric matrices in the replica space, with $\mathcal X_{ab}^{\sigma \sigma'} = \mathcal X_{ba}^{\sigma' \sigma}$, and explicitly
\[
    \mathcal D[\mathcal X]:=
    \prod_t
    \Bigl[\prod_{a\le b} d\mathcal X^{++}_{ab}(t)\,d\mathcal X^{--}_{ab}(t)\Bigr]
    \Bigl[\prod_{a,b} d\mathcal X^{+-}_{ab}(t)\Bigr],
\]
We thus arrive at the single-spin action conditioned to the field $\mathcal{X}$
\begin{equation} \label{app:S1-action}
    \mathcal{S}^{(1)}[\mathbf{s}; \mathcal{X}] =  \sum_{a, \sigma} \text{i} \sigma \mathcal{S}_{\rm top}[\mathbf{s}_{a,\sigma}] +  \frac{1}{2}   \sum_{ab,\sigma \sigma'} \int dt  \left( \gamma + \sqrt{-\sigma \sigma^{\prime}} \mathcal X^{\sigma \sigma'}_{a,b} \right)  \mathbf{s}_{a \sigma} \cdot \mathbf{s}_{b \sigma'}
    - 2 \gamma s(s+1) n T 
\end{equation}
in which the bilinears enter only through the matrix coupling between $\mathcal X^{\sigma\sigma'}_{ab}$ and $B^{\sigma\sigma'}_{ab}$.

\subsection{Equivalent Liouville-space derivation}
\label{app:liouville}

We now derive the same fixed-$\mathcal X$ single-site problem directly in Liouville space. We vectorize the replicated many-body density matrix using the standard convention
$\lket{A\rho B}=(A\otimes B^T)\lket{\rho}$, and define left/right spin superoperators by
\begin{equation}
\label{eq:leftrightspinsup}
    \hat S^\alpha_{i,a,+}\lket{\rho}\equiv\lket{\hat S_i^\alpha\rho},
    \qquad
    \hat S^\alpha_{i,a,-}\lket{\rho}\equiv\lket{\rho \hat S_i^\alpha}.
\end{equation}
Thus $\hat S^\alpha_{i,a,+}$ is represented by $\hat S_i^\alpha\otimes\mathbb 1$, while
$\hat S^\alpha_{i,a,-}$ is represented by $\mathbb 1\otimes(\hat S_i^\alpha)^T$.
For Hermitian spin operators this is the complex-conjugate representation, which is essential for the $y$ component.
After averaging the Brownian couplings and the unbiased measurement records, the non-normalized replicated many-body density matrix obeys
\begin{equation}
\label{app:manybody-liouville}
\begin{split}
    \partial_t\lket{\tilde\rho^{(n)}} =
    \Biggl[
    &\frac{\gamma}{2}
    \sum_{i,a,b,\sigma,\sigma'}
    \hat{\mathbf S}_{i,a,\sigma}\!\cdot\!\hat{\mathbf S}_{i,b,\sigma'}
    -\frac{J^2}{4N}
    \sum_{a,b,\sigma,\sigma'}
    \sigma\sigma'\,
    \bigl(\hat{\mathbb B}^{\sigma\sigma'}_{ab}\bigr)^2 \\
    &-2\gamma nNs(s+1)
    \Biggr]\lket{\tilde\rho^{(n)}} = \hat{\mathbb L}_{\rm MB}^{(n)} \lket{\tilde\rho^{(n)}},
    \qquad
    \hat{\mathbb B}^{\sigma\sigma'}_{ab}
    \equiv
    \sum_i
    \hat{\mathbf S}_{i,a,\sigma}\!\cdot\!\hat{\mathbf S}_{i,b,\sigma'} \;,
\end{split}
\end{equation}
where we use the subscript MB to refer to many-body quantities.
The fully connected structure is contained in the term quadratic in
$\hat{\mathbb B}$, with formally $\hat{\mathbb B}=O(N)$, as it is the sums over all spins. The natural mean-field variable is
therefore the intensive bilinear
$\hat{\mathbb B}^{\sigma\sigma'}_{ab}/N$, whose fluctuations are suppressed
at large $N$. Equivalently to the Hubbard-Stratonovich decoupling, one
expands around a self-consistent expectation value,
\[
    \frac{1}{N}\hat{\mathbb B}^{\sigma\sigma'}_{ab}
    =
    \frac{1}{N}\braket{\hat{\mathbb B}^{\sigma\sigma'}_{ab}}
    +\delta B^{\sigma\sigma'}_{ab},
    \qquad
    \delta B^{\sigma\sigma'}_{ab}
    =
    \frac{\hat{\mathbb B}^{\sigma\sigma'}_{ab}
    -\braket{\hat{\mathbb B}^{\sigma\sigma'}_{ab}}}{N},
\]
and keeps only terms linear in $\delta B$, an approximation that becomes exact at large $N$ as we will see. This gives
\begin{equation}
\label{eq:meanfieldexp}
    -\frac{\sigma\sigma'}{4N}
    \bigl(\hat{\mathbb B}^{\sigma\sigma'}_{ab}\bigr)^2
    \longrightarrow
    -\frac{\sigma\sigma'}{2N}
    \braket{\hat{\mathbb B}^{\sigma\sigma'}_{ab}}\,
    \hat{\mathbb B}^{\sigma\sigma'}_{ab}
    +\frac{\sigma\sigma'}{4N}
    \braket{\hat{\mathbb B}^{\sigma\sigma'}_{ab}}^2 .
\end{equation}
A clarification regarding the expectation value appearing in this last equation is useful. The time evolution in Eq.~\eqref{app:manybody-liouville} can be viewed as the application of an infinitesimal transfer matrix (generated by $\hat{\mathbb L}_{\rm MB}^{(n)}$) in the calculation of the partition function for the time trajectory of the $N$ spins. This sum is made explicit in the path integral version~\eqref{eq:pathintegralO}. In this interpretation, it is clear that the expectation value must be calculated over a specific time slice, with the boundary conditions determined by 1) the initial state $\lket{\rho_0} \propto \lket{\mathbb 1}$ and 2) the observable of interest, which fixes the boundary state $\lbra{\Omega^{(n)}}$. More explicitly
\begin{equation}
\label{eq:Bexpect}
    \braket{\hat{\mathbb B}^{\sigma\sigma'}_{ab}(t)}
    =
    \frac{\lbra{\Omega^{(n)}}\hat{\mathbb U}_{{\rm MB};T,t}^{(n)}\,
    \hat{\mathbb B}^{\sigma\sigma'}_{ab}\,
    \hat{\mathbb U}_{{\rm MB};t,0}^{(n)}\lket{\rho_0}}
    {\lbra{\Omega^{(n)}}\hat{\mathbb U}_{{\rm MB};T,0}^{(n)}\lket{\rho_0}},
\end{equation}
where $\hat{\mathbb U}_{{\rm MB};t_2,t_1}^{(n)} = e^{(t_2 - t_1)
\hat{\mathbb L}_{\rm MB}^{(n)}}$ is the many-body Liouville propagator
generated by~\eqref{app:manybody-liouville}.
At large $N$, the truncation in Eq.~\eqref{eq:meanfieldexp} becomes exact. For consistency with the previous section, we parametrize the expectation values as
\begin{equation}
\label{eq:XfromB}
    \frac{\sqrt{-\sigma\sigma'}}{N}\,
    \braket{\hat{\mathbb B}^{\sigma\sigma'}_{ab}(t)}
    = \mathcal X^{\sigma\sigma'}_{ab}(t),
    \qquad
    \frac{1}{N}\,
    \braket{\hat{\mathbb B}^{\sigma\sigma'}_{ab}(t)}
    =: X^{\sigma\sigma'}_{ab}(t).
\end{equation}
Here $\mathcal X$ is consistent with the Hubbard-Stratonovich field introduced in the previous section, while $X$
is obtained after a rotation in the complex plane, with
\[
    X^{+-}_{ab}=\mathcal X^{+-}_{ab},
    \qquad
    X^{++}_{ab}=-i\mathcal X^{++}_{ab},
    \qquad
    X^{--}_{ab}=-i\mathcal X^{--}_{ab}.
\]
In the remainder of this section we use the rotated variables $X$.
Consequently, at fixed $X(t)$ the conditioned many-body generator separates into a scalar contribution and a sum of identical single-site generators,
\[
    \hat{\mathbb L}_{\rm MB}^{(n)}(t) \longrightarrow \hat{\mathbb L}_{{\rm MB};X}^{(n)}(t)
    =
    \frac{N}{4}\sum_{a,b,\sigma,\sigma'}
    \sigma\sigma' \bigl(X^{\sigma\sigma'}_{ab}(t)\bigr)^2
    +
    \sum_{i=1}^N\hat{\mathbb L}^{(n)}_{X,i}(t).
\]
The resulting
single-site Liouville generator is the operator counterpart of
Eq.~\eqref{app:S1-action} and reads
\begin{equation}
    \hat{\mathbb L}_{X}^{(n)}(t):=
    \frac{1}{2}
    \sum_{a,b,\sigma,\sigma'}
    \left(\gamma-\sigma\sigma' X^{\sigma\sigma'}_{ab}(t)\right)
    \hat{\mathbf S}_{a,\sigma}\!\cdot\!\hat{\mathbf S}_{b,\sigma'}
    -2\gamma ns(s+1).
\end{equation}
Note that the quadratic term in $X$ is included in the conditioned many-body generator $\hat{\mathbb L}_{{\rm MB};X}^{(n)}$, but not in the single-site generator $\hat{\mathbb L}_{X}^{(n)}$.
When the time evolution in Eq.~\eqref{eq:Bexpect} is replaced by that for decoupled spins for a given field $X(t)$, Eqs.~\eqref{eq:Bexpect} and \eqref{eq:XfromB} become self-consistent conditions determining $X(t)$:
\begin{equation}
\label{eq:Xselfconst}
    X^{\sigma\sigma'}_{ab}(t)
    =
    \frac{\lbra{\Omega^{(n)}}\hat{\mathbb U}_{X;T,t}^{(n)}\,
    \hat{\mathbf S}_{a,\sigma}\!\cdot\!\hat{\mathbf S}_{b,\sigma'} \,
    \hat{\mathbb U}_{X;t,0}^{(n)}\lket{\rho_0}}
    {\lbra{\Omega^{(n)}}\hat{\mathbb U}_{X;T,0}^{(n)}\lket{\rho_0}} =: F_{ab}^{\sigma \sigma'}[X](t).
\end{equation}
As expected, it is easy to verify that this condition is precisely the saddle point condition dominating the path integral in Eq.~\eqref{eq:EOpathint} in the limit of large $N$. In this equation we have removed the lattice index $i$, as the dynamics only concerns one single spin. It is also useful to present the \textit{unvectorized} form of these equations. Here $\mathcal U_{X;t_2,t_1}$ denotes the corresponding non-vectorized evolution map on density matrices, defined by
\[
    \lket{\mathcal U_{X;t_2,t_1}[\rho]}
    =
    \hat{\mathbb U}_{X;t_2,t_1}^{(n)}\lket{\rho}.
\]
\begin{equation}
    X^{+-}_{ab}(t)=
    \frac{\Tr\!\left[\Omega^{(n)}\, \mathcal U_{X;t,T}
    \Bigl[\sum_\alpha \hat S_a^\alpha \trho1^{(n)}(t) \hat S_b^\alpha\Bigr]\right]}
    {\Tr[\Omega^{(n)}\,\trho1^{(n)}(T)]},
\end{equation}
\begin{equation}
    X^{++}_{ab}(t)=
    \frac{\Tr\!\left[\Omega^{(n)}\, \mathcal U_{X;t,T}
    \Bigl[(\hat{\mathbf S}_a\!\cdot\!\hat{\mathbf S}_b)\trho1^{(n)}(t)\Bigr]\right]}
    {\Tr[\Omega^{(n)}\,\trho1^{(n)}(T)]},
\end{equation}
\begin{equation}
    X^{--}_{ab}(t)=
    \frac{\Tr\!\left[\Omega^{(n)}\, \mathcal U_{X;t,T}
    \Bigl[\trho1^{(n)}(t)(\hat{\mathbf S}_a\!\cdot\!\hat{\mathbf S}_b)\Bigr]\right]}
    {\Tr[\Omega^{(n)}\,\trho1^{(n)}(T)]}.
\end{equation}
so that $\trho1^{(n)}(t)=\mathcal U_{X;0,t}[\trho1(0)]$.
The unreduced single-site evolution in unvectorized form has the form
\begin{equation}
\label{app:mathLn-unred}
\begin{split}
    \frac{d}{dt}\trho1^{(n)} =
    \mathcal L_X^{(n)}(\trho1^{(n)})
    =
    \sum_{ab} \left[ (\gamma + X_{ab}^{+-}) \hat S^{\beta}_{a}\trho1^{(n)}  \hat S^{\beta}_{b} +  \frac{1}{2} (\gamma - X_{a,b}^{++}) (\hat{\mathbf S}_{a} \cdot \hat{\mathbf S}_{b})\trho1^{(n)} + \frac{1}{2} (\gamma - X_{a,b}^{--} )\trho1^{(n)} (\hat{\mathbf S}_{a} \cdot \hat{\mathbf S}_{b}) \right]
    -  2 \gamma n s (s+1)\trho1^{(n)} .
\end{split}
\end{equation}

\subsection{Symmetries of the replicated dynamics}
\label{app:unreduced-symmetry}

The symmetry structure relevant for the saddle-point Ansatz is most transparent
before any mean-field reduction. At the many-body level, the replicated
Liouville generator appearing in Eq.~\eqref{app:manybody-liouville} is invariant under
independent permutations of the replicas on the left and right density-matrix
indices, namely under a pair of permutations $(P_+,P_-) \in S_n^+\times S_n^-$. In the vectorized form, we set
\[
    \hat{\mathbb U}_{P_+,P_-}\lket{\rho}
    \equiv
    \lket{U_{P_+}\rho\,U_{P_-}^{-1}},
\]
where $U_{P_+}$ and $U_{P_-}$ act on the $+$ and $-$ replica indices,
respectively. Then
\[
    \hat{\mathbb U}_{P_+,P_-}\,
    \hat{\mathbf S}_{i,a,\sigma}\,
    \hat{\mathbb U}_{P_+,P_-}^{-1}
    =
    \hat{\mathbf S}_{i,P_\sigma(a),\sigma},
\]
and therefore
\[
    \hat{\mathbb U}_{P_+,P_-}\,
    \hat{\mathbb B}_{ab}^{\sigma\sigma'}\,
    \hat{\mathbb U}_{P_+,P_-}^{-1}
    =
    \hat{\mathbb B}_{P_\sigma(a)\,P_{\sigma'}(b)}^{\sigma\sigma'}.
\]
Since Eq.~\eqref{app:manybody-liouville} is summed over all replica labels, the many-body generator is
invariant under this action of $S_n\times S_n$.

There is also an intrinsic contour-exchange $\mathbb Z_2$, realized as the
antilinear map
\[
    \mathcal C_{\rm MB}\lket{\rho}\equiv\lket{\rho^\dagger}.
\]
On elementary superoperators it acts as
\[
    \mathcal C_{\rm MB}\,\hat{\mathbf S}_{i,a,+}\,\mathcal C_{\rm MB}^{-1}
    =
    \hat{\mathbf S}_{i,a,-},
    \qquad
    \mathcal C_{\rm MB}\,\hat{\mathbf S}_{i,a,-}\,\mathcal C_{\rm MB}^{-1}
    =
    \hat{\mathbf S}_{i,a,+}.
\]
Note that $\mathcal C_{\rm MB}$ is antilinear and that these relations rely on the Hermiticity of the spin operators: e.g.\
$\mathcal C_{\rm MB}\,\hat{S}^\alpha_{i,a,+}\,\mathcal C_{\rm MB}^{-1}\lket{\rho}
= \mathcal C_{\rm MB}\lket{\hat{S}^\alpha_{i,a}\, \rho^\dagger}
= \lket{\rho\, (\hat{S}^{\alpha}_{i,a})^\dagger}
= \lket{\rho\, \hat{S}^\alpha_{i,a}}$,
in agreement with the definition \eqref{eq:leftrightspinsup} of the right action. The antilinearity of $\mathcal C_{\rm MB}$ is also responsible for the complex conjugation appearing in Eq.~\eqref{eq:contour} below. Therefore
\[
    \mathcal C_{\rm MB}\,
    \hat{\mathbb B}_{ab}^{\sigma\sigma'}\,
    \mathcal C_{\rm MB}^{-1}
    =
    \hat{\mathbb B}_{ab}^{-\sigma,-\sigma'}.
\]
Because the coefficients of Eq.~\eqref{app:manybody-liouville} are real and depend on
$\sigma,\sigma'$ only through the combination $\sigma\sigma'$, the many-body
generator is covariant under $\mathcal C_{\rm MB}$. Altogether, the unreduced
replicated problem has symmetry $(S_n\times S_n)\rtimes\mathbb Z_2$.

These symmetries are inherited by the mean-field fields, as follows directly from the self-consistency equations \eqref{eq:Xselfconst}. More explicitly, if $X_{ab}^{\sigma \sigma'}(t)$ is a solution, then $X_{P_\sigma(a)\,P_{\sigma'}(b)}^{\sigma\sigma'}(t)$ is also a solution for any $(P_+, P_-) \in S_n \times S_n$.

For the contour exchange, let us set
\begin{equation}
\label{eq:contour}
(\mathcal{C}X)_{ab}^{\sigma \sigma'} :=
\bigl(X_{ab}^{-\sigma, -\sigma'}\bigr)^\ast .
\end{equation}
Then, using that $\hat{\mathcal C}_{\rm MB}^2 = 1$ and that
\[
    \hat{\mathcal C}_{\rm MB}\hat{\mathbb U}_{X;t_2,t_1}^{(n)}\hat{\mathcal C}_{\rm MB}
    =
    \hat{\mathbb U}_{\mathcal C X;t_2,t_1}^{(n)},
\]
we deduce, for $\mathcal C$-invariant boundary states, namely
$\hat{\mathcal C}_{\rm MB}\lket{\mathbb 1}=\lket{\mathbb 1}$ and
$\lbra{\Omega^{(n)}}\hat{\mathcal C}_{\rm MB}=\lbra{\Omega^{(n)}}$, that the expectation value in Eq.~\eqref{eq:Xselfconst} satisfies
\begin{equation}
 [F_{ab}^{\sigma \sigma'}[X](t)]^\ast = F_{ab}^{-\sigma -\sigma'}[\mathcal C X](t).
\end{equation}
This is the local contour-exchange covariance of the self-consistency map. Under this additional
assumption, if $X(t)$ is a solution, then so is $\mathcal C X(t)$.

Moreover, there is an extra symmetry associated to time reversal. Explicitly,
\begin{equation}
 [F_{ab}^{\sigma \sigma'}[X](t)]^\ast =
 \frac{\lbra{\rho_0}\hat{\mathbb U}_{X;T,t}^{(n)\dagger}\,
    \hat{\mathbf S}_{a,\sigma}\!\cdot\!\hat{\mathbf S}_{b,\sigma'} \,
    \hat{\mathbb U}_{X;t,0}^{(n) \dagger}\lket{\rho_0}}
    {\lbra{\rho_0}\hat{\mathbb U}_{X;T,0}^{(n)}\lket{\Omega^{(n)}}} \;.
\end{equation}
Thus, using that $\hat{\mathbb U}_{X; t_2, t_1}^{(n)\dagger} = \hat{\mathbb U}_{X^\ast; t_1, t_2}^{(n)} = \hat{\mathbb U}_{\mathcal{T} X; t_2, t_1}^{(n)}$, with $\mathcal{T}X(t) = X(T-t)^\ast$, we deduce that for $\rho_0 = \mathbb 1/2$ and $\Omega^{(n)} = \mathbb 1$,
\begin{equation}
 [F_{ab}^{\sigma \sigma'}[X](t)]^\ast =  [F_{ab}^{\sigma \sigma'}[\mathcal{T} X](T-t)]
\end{equation}
so that if $X(t)$ is a solution, $\mathcal{T}X(t)$ is also a solution of the saddle point equation.

In the disordered phase (strong measures), it is expected that none of these symmetries will be broken; in other words, there is a single physical solution to the saddle point equation. In the bulk $0 \ll t \ll T$, it becomes time independent with the form
\begin{equation} \label{app:ansatzX}
    \begin{split}
        \bar{X}^{+-}_{a,b} &= \bar{X}^{-+}_{a,b}  = r \\
        \bar{X}^{++}_{a,b} &= \bar{X}^{--}_{a,b} = q (1-\delta_{a,b}) + s(s+1) \delta_{a,b},
    \end{split}
\end{equation}
In general, we assume that the contour-exchange symmetry \eqref{eq:contour} is never broken. Furthermore, we expect $X$ to remain real at all times, as justified in the bulk by the time-reversal symmetry presented earlier. These two conditions ensure that for every $t \in [0, T]$, one must have 
\begin {equation}
\label{eq:reducedX}
X^{++}_{ab}(t) = X^{--}_{ab}(t) \;, \qquad X^{+-}_{ab}(t) = X^{-+}_{ab}(t) = X^{+-}_{ba}(t) \;.
\end{equation}
Quantitatively, the effective action in Eq.~\eqref{eq:EOpathint}
takes the form
\begin{multline} \label{mathI}
    \mathcal{I} [X] =
    -\frac{1}{T}\ln \tr\bigl[\Omega \trho1(T)\bigr]
    -\frac{1}{4 T}\int dt
    \sum_{a,b,\sigma,\sigma'}
    \sigma\sigma' \bigl(X^{\sigma\sigma'}_{ab}(t)\bigr)^2
     =\\=
    \frac{1}{2T} \int dt \sum_{a,b} \left((X^{+-}_{a,b})^2 - (X^{++}_{a,b})^2\right) - \frac{1}{T} \ln \operatorname{tr} [\Omega \trho1 (T)]
\end{multline}
where in the second equality we used \eqref{eq:reducedX} and
\begin{equation}
\begin{split} \label{mathLn}
    \frac{d}{dt}\trho1^{(n)} \equiv \mathcal{L}_X^{(n)}(\trho1^{(n)}) =  \sum_{ab} \left[ (\gamma + X_{ab}^{+-} (t)) \hat S^{\alpha}_{a}\trho1^{(n)}  \hat S^{\alpha}_{b} + \frac{1}{2} (\gamma - X_{a,b}^{++}(t)) \lbrace \hat{\mathbf S}_{a} \cdot \hat{\mathbf S}_{b},\trho1^{(n)} \rbrace  \right]  - 2 \gamma n s (s+1)\trho1^{(n)} .
\end{split}
\end{equation}
For $s=1/2$, this reduces to Eqs.~\eqref{eq:Idefmain} and \eqref{main:mathLn} of the main text. 

Consistently with the scenario of purely unitary dynamics ($\gamma=0$), in the weak-measurement phase, the symmetry $S_n^+ \times S_n^-$ is expected to break down into a single $S_n$, with a residual $S_n \simeq S_n \times S_n / S_n$. Specifically, there are several solutions identified by the sector $P \in S_n$, with a residual symmetry $P_- \in S_n$ and
\begin{equation}
    P_+=P\,P_-\,P^{-1}.
\end{equation}
To explore the possibility of a symmetry breaking, we can select a particular sector $P=\mathbb 1$ (as done in the main text), leading to the time-dependent Ansatz
\begin{equation} \label{app:ansatzXtime}
    \begin{split}
        \bar{X}^{+-}_{a,b}(t)&= r(t) + X(t) \delta_{ab} \\
        \bar{X}^{++}_{a,b}(t) &= 
        q(t) (1-\delta_{a,b}) + s(s+1) \delta_{a,b},
    \end{split}
\end{equation}
The corresponding solutions in the other sectors are generated by acting with $P$ on the $+$ replica indices of the mixed-contour field only,
\begin{equation}
    X^{+-}_{ab}[P](t)=\bar X^{+-}_{P(a),b}(t),
\end{equation}
while $\bar X^{++}$ is unchanged because of the replica-symmetric form of the identity-sector Ansatz. Equivalently, in the reduced evolution \eqref{mathLn}, the sector $P$ is obtained by permuting only the replica labels carried by the operators acting on the left of $\trho1$, namely $\hat S_a^\alpha \mapsto \hat S_{P(a)}^\alpha$, while the right labels are left untouched. At the level of the reduced density matrix this corresponds to a left action only,
\begin{equation}
\label{eq:othersectors}
    \trho1^{P}(t)=\hat R_{+}[P]\,\trho1^{\mathbb 1}(t) \;.
\end{equation}

\section{Saddle-point solution}

\subsection{Derivation of the effective action}\label{app:LGeff}
\noindent
Starting from the reduced action \eqref{mathI},
we insert the Ansatz \eqref{app:ansatzX}, focusing on the bulk where all parameters are assumed to be time-independent and we focus on the large $T$ limit. The quadratic part is then evaluated exactly as
\begin{equation}
    \sum_{a,b}(\bar X^{+-}_{ab})^2
    =
    n(n-1)r^2+n(r+X)^2 \;, \qquad
    \sum_{a,b}(\bar X^{++}_{ab})^2
    =
    n(n-1)q^2+n\left(\frac34\right)^2.
\end{equation}
To treat the logarithmic term, we observe that at large $T$, the evolution \eqref{mathLn}, with the Ansatz \eqref{app:ansatzX}, will project onto the largest eigenvalue $\lambda_n$ of the static generator $\mathcal{L}_X^{(n)}$. After inserting the static Ansatz, this generator takes the explicit form
\begin{equation}
    \mathcal L_{\bar X}^{(n)}(\trho1^{(n)})
    =
    (\gamma+r)\sum_\alpha \hat S^\alpha_{\rm tot}\trho1^{(n)} \hat S^\alpha_{\rm tot}
    +\frac12(\gamma-q)\{\hat{\mathbf S}_{\rm tot}^2,\trho1^{(n)}\}
    +X\sum_{a,\alpha}\hat S_a^\alpha \trho1^{(n)} \hat S_a^\alpha
    +\frac{3n}{4}\left(q-2\gamma-\frac34\right)\trho1^{(n)},
\end{equation}
so it preserves the sectors of fixed total spin $\hat{\mathbf S}_{\rm tot}^2$, where
\begin{equation}
 \hat{\mathbf S}_{\rm tot} = \sum_{a} \hat{\mathbf{S}}_a
\end{equation}
is the sum over all replicas, on a given site.  The eigenspace with total spin
\(\ell\) is not, in general, a single irreducible representation: it is the
direct sum of \(d_\ell^{(n)}\) equivalent copies of the spin-\(\ell\) irrep (see Sec.~\ref{sec:derBmatr}).
We denote by \(P_\ell\) the projector onto this full isotypic component and use
the Hilbert-Schmidt normalized operator
$\Pi_\ell=P_\ell/\sqrt{D_{n,\ell}}$, with
$D_{n,\ell}=\Tr P_\ell=(2\ell+1)d_\ell^{(n)} = \Tr[\Pi_\ell]^2$.  Thus
\(\Tr(\Pi_\ell\Pi_{\ell'})=\delta_{\ell,\ell'}\).
As explained in the main text, the evolution through $\mathcal{L}_{\bar{X}}^{(n)}$ preserves the linear combinations of $\Pi_{\ell}$, which includes the initial condition
\begin{equation}
 \tilde\varrho^{(n)}_0 = \mathbb{1}/2 =\frac12 \sum_{\ell} P_\ell = \frac12 \sum_{\ell} \sqrt{D_{n,\ell}} \Pi_\ell \;.
\end{equation}
So, we can look for the leading eigenstate of  the matrix relating different sectors of $\ell$
\begin{equation}
\label{eq:Lproj}
    (\mathcal L_X^{(n)})_{\ell,\ell'}
    :=
    \Tr\!\left[\Pi_{\ell'}\,\mathcal L_X^{(n)}(\Pi_\ell)\right]
    =
    (2\gamma+r-q)L_{\ell,\ell'}
    +\frac{3n}{4}\left(q-2\gamma-\frac34\right)\delta_{\ell,\ell'},
\end{equation}
where $\ell$ runs over the total-spin values obtained by adding $n$ spin-$1/2$ replicas, namely $\ell=0,1,\ldots,n/2$ for even $n$ and $\ell=\frac12,\frac32,\ldots,\frac{n}{2}$ for odd $n$. Therefore $L_{\ell,\ell'}$ has dimension $\lfloor n/2\rfloor+1$.
Its explicit form can be obtained using that $[\hat S_{\rm tot}^\alpha, \Pi_\ell] = 0$, which leads to
\begin{equation}
\label{eq:LdefwithB}
    L_{\ell,\ell'}
    =
    \ell(\ell+1)\delta_{\ell,\ell'}
    +x\sum_{a,\alpha}\Tr\!\left[
    \Pi_\ell \hat S_a^\alpha \Pi_{\ell'} \hat S_a^\alpha
    \right] = \ell(\ell+1)\delta_{\ell,\ell'}
    +x B_{\ell, \ell'},
    \qquad
    x=\frac{X}{2\gamma+r-q}.
\end{equation}
where we implicitly defined the matrix $B$. We postpone the technical calculation of the matrix elements $B_{\ell, \ell'}$ in the next subsection, but its tridiagonal structure could be understood as a consequence of the compositions of $SU(2)$ irreps: $S_a^\alpha \Pi_{\ell'} S_a^\alpha$ corresponds to a spin $\ell'$ together with two spin $1/2$. Furthermore, Eqs.~(\ref{eq:Lproj},\ref{eq:LdefwithB}) provides a useful way to separate the dependence on the three parameters $(X, q, r)$ of the Ansatz \eqref{app:ansatzX}. If $\lambda_n(X,q,r)$ denotes the maximal eigenvalue of the bulk generator $\mathcal L_X^{(n)}$ and $\Lambda_n(x)$ the maximal eigenvalue of $L$, then
\begin{equation}
    \lambda_n(X,q,r)
    =
    (2\gamma+r-q)\Lambda_n(x)
    +\frac{3n}{4}\left(q-2\gamma-\frac34\right),
\end{equation}
and the complete static action takes the form
\begin{equation}
    \mathcal I_{\rm full}(q,r,X)
    =
    \frac{n(n-1)}{2}(r^2-q^2)
    +\frac{n}{2}(r+X)^2
    -\frac{9n}{32}
    -\lambda_n(X,q,r).
\end{equation}
The extra constant $-9n/32$ comes from the diagonal constraint
$X^{++}_{aa}=s(s+1)=3/4$ and is needed for the correct normalization at
$n=1$.

The saddle equations are now simply obtained differentiating with respect to the three parameters $q ,r, X$, leading to
\begin{align}
\label{eq:saddleqrX}
    n(n-1)q-\Lambda_n(x)+x\Lambda_n'(x)+\frac{3}{4}n&=0,\nonumber\\
    n(nr+X)-\Lambda_n(x)+x\Lambda_n'(x)&=0,\nonumber\\
    n(r+X)-\Lambda_n'(x)&=0,
\end{align}
with $x=X/(2\gamma+r-q)$. Eliminating $q$, $r$, and $\Lambda_n'(x)$ gives the exact one-parameter action
\begin{equation}
    \mathcal I^{\rm full}_n(x)
    =
    \frac{\left[n(n+x)\left(\frac34+2(n-1)\gamma\right)-\Lambda_n(x)\right]^2}
    {2n(n-1)(n+2x)}
    -\frac{\gamma n(n-1)}{2}(4\gamma n+3).
\end{equation}
Therefore the compact formula used in the main text,
\begin{equation}
    \mathcal I_n(x)
    =
    \frac{\left[n(n+x)\left(\frac34+2(n-1)\gamma\right)-\Lambda_n(x)\right]^2}
    {2n(n-1)(n+2x)},
\end{equation}
differs from the fully normalized one only by an $x$-independent term vanishing at $n = 1$.

As a consistency check, setting $n = 1$, one has $\Lambda_1(x)=\frac34(1+x)$ and one finds $\lim_{n\to1}\mathcal I^{\rm full}_n(x)=0$ for every $x$, consistently with the preservation of the trace of the standard Lindbladian generator ($n=1$).

\subsubsection{Derivation of the matrix representation \label{sec:derBmatr}}
\noindent
We now derive the explicit form of the matrix $L$ in \eqref{eq:LdefwithB}, specializing to the case of $s=1/2$.  The Hilbert space decomposes into isotypic irreps of ${\rm SU}(2)$ as
\[
    \left(\mathbb C^2\right)^{\otimes n}
    =
    \bigoplus_\ell
    \left(V_\ell\otimes \mathbb C^{d_\ell^{(n)}}\right),
\]
where \(V_\ell\) is the spin-\(\ell\) irrep of \(SU(2)\) and
\(d_\ell^{(n)}\) is its multiplicity.  Therefore the dimension of the full
eigenspace of \(S^2_{\rm tot}\) is
\begin{equation}
    D_{n,\ell} = (2 \ell + 1) d^{(n)}_\ell
\end{equation}
where $d^{(n)}_\ell$ is the generalized Catalan number 
\begin{equation}
d^{(n)}_\ell \equiv C_{\ell + n/2,n/2- \ell} = \frac{2 \ell + 1}{n+1} \begin{pmatrix}
n +1 \\ n/2 - \ell
\end{pmatrix}
\end{equation}
which accounts for the multiplicity of the spin-\(\ell\) representation in the
addition of \(n\) spin \(1/2\)'s.  Let
\(\ket{\ell,M,\mu}\), with \(\mu=1,\ldots,d_\ell^{(n)}\), be an orthonormal
basis of the full isotypic component.  We will see what is a convenient choice for the index $\mu$. The operator used in the matrix
representation is the Hilbert-Schmidt normalized isotypic projector
\begin{equation}
\label{eq:Piell-isotypic}
    \Pi_{\ell}
    =
    \frac{1}{\sqrt{D_{n,\ell}}}
    \sum_{\mu=1}^{d_\ell^{(n)}}
    \sum_{M=-\ell}^{\ell}
    \ket{\ell,M,\mu}\bra{\ell,M,\mu}.
\end{equation}
Because of the permutational and rotational invariance:
\begin{equation} \label{app:matrixB}
   B_{\ell, \ell'} \equiv \sum_{a,\alpha} \Tr [\Pi_{\ell} \hat S_a^{\alpha} \Pi_{\ell'} \hat S^{\alpha}_a] = 3n \Tr [\Pi_{\ell} \hat S_1^{z} \Pi_{\ell'} \hat S^{z}_1]  . 
\end{equation}
Let us now choose as a basis the tensor product of the single spin $\hat S_1^z$ and of the sum of the other $n-1$ ones. The Hilbert space corresponding to the latter can be decomposed into a collection of spin-$j$ representations, each generated by $\ket{j,m,\nu}$, with $m=-j,\ldots,j$ and
\(\nu=1,\ldots,d_j^{(n-1)}\).  Coupling this spin \(j\) to the last spin
\(1/2\), we use a basis
\(\ket{\ell,M;j,\nu}\) for the spin-\(\ell\) isotypic component.  The label
\((j,\nu)\) is a valid multiplicity label $\mu$ (see Eq.~\eqref{eq:Piell-isotypic}) for the spin-\(\ell\) sector whenever
\(\ell=j\pm1/2\).  In this coupled basis,
\begin{equation}
\label{eq:Piell-coupled}
    \Pi_\ell
    =
    \frac{1}{\sqrt{D_{n,\ell}}}
    \sum_{\substack{j,\nu:\\ \ell=j\pm1/2}}
    \sum_{M=-\ell}^{\ell}
    \ket{\ell,M;\mu \equiv (j,\nu)}\bra{\ell,M;\mu \equiv (j,\nu)}.
\end{equation}
This is the same projector as in Eq.~\eqref{eq:Piell-isotypic}, but with a choice of the index $\mu$ adapted to the $(n-1)+1$ decomposition.
The uncoupled basis obtained fusing the $n-1$ spins with the last spin $1/2$
\begin{equation}
\left\{\ket{j, m,\nu, \sigma} \equiv \ket{j, m,\nu}\otimes \ket{\frac{1}{2}, \frac{\sigma}{2}} \right\}_{j, \nu, \sigma = \pm 1}
\end{equation}
constitute a basis for the whole $n$-replica Hilbert space.

As is often the case in this type of calculation, some quantities are best expressed in the uncoupled basis, while others are best expressed in the coupled basis. The relationship between the two is obtained using the Clebsch-Gordan coefficients. Explicitly, for the matrix element of $\hat S_1^z$, we simply have
\begin{equation}
\label{app:Szmatrel}
    \bra{j', m',\nu', \sigma'} \hat S_1^z \ket{j, m,\nu, \sigma} = \frac{\sigma}{2} \delta_{jj'} \delta_{m m'} \delta_{\nu\nu'}\delta_{\sigma \sigma'} \;.
\end{equation}
The matrix elements of \(\Pi_\ell\) in the uncoupled basis involve the Clebsch-Gordan coefficients
\begin{equation}
\label{eq:Piell-matrix-multiplicity}
\bra{j',m',\nu',\sigma'}\Pi_\ell
\ket{j,m,\nu,\sigma}
=
\frac{\delta_{jj'}\delta_{\nu\nu'}}
{\sqrt{D_{n,\ell}}}
\sum_{M=-\ell}^{\ell}
\braket{j,m,\sigma|\ell M}
\braket{\ell M|j,m',\sigma'} .
\end{equation}
Note that the diagonality in \(j, j'\) and \(\nu, \nu'\) follows automatically from the convenient choice of coupled basis \(\ket{\ell,M;\mu \equiv (j,\nu)}\). In the last line of Eq.~\eqref{eq:Piell-matrix-multiplicity} we omit the multiplicity index as it is now inessential.

Inserting Eqs.~(\ref{eq:Piell-matrix-multiplicity}, \ref{app:Szmatrel}) in Eq.~\eqref{app:matrixB} gives
\begin{equation}
\begin{split}
    B_{\ell, \ell'} &= \frac{3}{4}  \frac{n}{\sqrt{D_{n,\ell} D_{n,\ell'}}} \sum_{j,m,m',\sigma,\sigma'} \sigma \sigma^{\prime} d^{(n-1)}_j \bra{j, m, \sigma} \Pi_{\ell} \ket{j, m', \sigma'} \bra{j, m', \sigma'} \Pi_{\ell'} \ket{j, m, \sigma} \\
    &= \frac{3}{4}  \frac{n}{\sqrt{D_{n,\ell} D_{n,\ell'}}} \sum \sigma \sigma^{\prime} d^{(n-1)}_j \braket{j, m, \sigma | \ell M} \braket{\ell M | j, m', \sigma'} \braket{j, m', \sigma'| \ell^{\prime} M'} \braket{\ell^{\prime} M' | j, m, \sigma} .
\end{split}
\end{equation}
where the sum over $\nu$ simply gives the multiplicity factor $d_j^{(n-1)}$. Of the above Clebsch-Gordan coefficients, the only nonzero ones are
\begin{equation}
\begin{split}
    \braket{ j = \ell + \scriptstyle \frac{1}{2} \displaystyle, m, \sigma | \ell M} &= \sqrt{\frac{1}{2} \left( 1 + \frac{ \sigma M}{\ell + 1} \right)} \delta_{M, m +  \scriptstyle \frac{\sigma}{2} \displaystyle} \\
    \braket{j = \ell - \scriptstyle \frac{1}{2} \displaystyle, m, \sigma | \ell M} &= - \sigma \sqrt{\frac{1}{2} \left( 1 - \frac{\sigma M}{\ell} \right)} \delta_{M, m +  \scriptstyle \frac{\sigma}{2} \displaystyle} . 
\end{split}
\end{equation}
This implies that the matrix elements vanish for $|\ell - \ell^{\prime}| > 1$, so $L$ is tridiagonal as anticipated. For $\ell^{\prime} = \ell + 1$, the only contribution of the sum over $j$ comes from $j = \ell + 1/2 = \ell' - 1/2$, i.e.
\begin{equation}
\begin{split}
    B_{\ell,\ell+1} &= \frac{3}{4} \frac{n d^{(n-1)}_{\ell + \scriptstyle \frac{1}{2} \displaystyle}}{\sqrt{D_{n,\ell} D_{n,\ell+1}}} \sum \sigma \sigma^{\prime} \braket{\ell + \scriptstyle \frac{1}{2} \displaystyle, m, \sigma | \ell M} \braket{\ell M | \ell + \scriptstyle \frac{1}{2} \displaystyle , m', \sigma'} \braket{\ell + \scriptstyle \frac{1}{2} \displaystyle, m', \sigma'| \ell + 1, M'} \braket{\ell + 1, M' | \ell + \scriptstyle \frac{1}{2} \displaystyle, m, \sigma} \\
     & = \frac{3}{16} \frac{n d^{(n-1)}_{\ell + \scriptstyle \frac{1}{2} \displaystyle}}{\sqrt{D_{n,\ell} D_{n,\ell+1}}} \sum \delta_{M, m + \scriptstyle \frac{\sigma}{2} \displaystyle} \delta_{M', m' + \scriptstyle \frac{\sigma'}{2} \displaystyle} \delta_{M'M} \sigma^2 \sigma^{\prime 2} \sqrt{\left( 1 + \frac{\sigma M}{\ell +1} \right) \left(1 + \frac{\sigma' M}{\ell +1} \right) \left( 1 - \frac{\sigma' M'}{\ell +1} \right) \left( 1 - \frac{\sigma M'}{\ell +1} \right)} \\
     & = \frac{3}{4} \frac{n d^{(n-1)}_{\ell + \scriptstyle \frac{1}{2} \displaystyle}}{\sqrt{D_{n,\ell} D_{n,\ell+1}}} \sum_{M=-\ell}^{\ell} \left(1 - \frac{M^2}{(\ell +1)^2} \right) \\
     &= \frac{3}{4} \frac{n d^{(n-1)}_{\ell + \scriptstyle \frac{1}{2} \displaystyle}}{\sqrt{D_{n,\ell} D_{n,\ell+1}}}  (2 \ell +1) \left(1 - \frac{\ell}{3(\ell +1)} \right) \\
      &= \frac{n}{4} \frac{d^{(n-1)}_{\ell + \scriptstyle \frac{1}{2} \displaystyle}}{ \sqrt{D_{n,\ell} D_{n,\ell+1}}} \frac{(2 \ell +1)(2 \ell + 3)}{\ell +1}\\
    &= \frac{1}{2} \sqrt{\left( \frac{n}{2} - \ell \right) \left( \frac{n}{2} + \ell  + 2 \right) }
   \end{split}
\end{equation}
Let us consider now the case $\ell = \ell^{\prime}$: in this case, for $\ell \neq 0$, we have both the contribution $j = \ell - 1/2 $ and $j = \ell + 1/2$
\begin{equation}
\begin{split}
    B_{\ell,\ell} &= \frac{3}{4 D_{n,\ell}} n d^{(n-1)}_{\ell+ \scriptstyle \frac{1}{2} \displaystyle} \sum \sigma \sigma^{\prime}  \braket{\ell+ \scriptstyle \frac{1}{2} \displaystyle, m, \sigma | \ell M} \braket{\ell M | \ell+ \scriptstyle \frac{1}{2} \displaystyle, m', \sigma'} \braket{\ell+ \scriptstyle \frac{1}{2} \displaystyle, m', \sigma'| \ell M'} \braket{\ell M' | \ell+ \scriptstyle \frac{1}{2} \displaystyle, m, \sigma} \\
    &+ \frac{3}{4 D_{n,\ell}} n d^{(n-1)}_{\ell- \scriptstyle \frac{1}{2} \displaystyle} \sum \sigma \sigma^{\prime}  \braket{\ell- \scriptstyle \frac{1}{2} \displaystyle, m, \sigma | \ell M} \braket{\ell M | \ell-\scriptstyle \frac{1}{2} \displaystyle, m', \sigma'} \braket{\ell- \scriptstyle \frac{1}{2} \displaystyle, m', \sigma'| \ell M'} \braket{\ell M' | \ell- \scriptstyle \frac{1}{2} \displaystyle, m, \sigma} \\
    &= \frac{3}{16 D_{n,\ell}} n d^{(n-1)}_{\ell+ \scriptstyle \frac{1}{2} \displaystyle} \sum \sigma \sigma^{\prime} \delta_{M, m + \scriptstyle \frac{\sigma}{2} \displaystyle} \delta_{M', m' + \scriptstyle \frac{\sigma'}{2} \displaystyle} \delta_{M'M} \left( 1 + \frac{\sigma M}{\ell + 1} \right) \left( 1 + \frac{\sigma' M}{\ell + 1} \right)  \\
    &+ \frac{3}{16 D_{n,\ell}} n d^{(n-1)}_{\ell- \scriptstyle \frac{1}{2} \displaystyle} \sum 
    \sigma \sigma^{\prime} \delta_{M, m + \scriptstyle \frac{\sigma}{2} \displaystyle} \delta_{M', m' + \scriptstyle \frac{\sigma'}{2} \displaystyle} \delta_{M'M} \left( 1 - \frac{\sigma M}{\ell} \right) \left( 1 - \frac{\sigma' M}{\ell} \right) \\
    &= \frac{3n}{4D_{n,\ell}} \left( \frac{d^{(n-1)}_{\ell+ \scriptstyle \frac{1}{2} \displaystyle}}{(\ell + 1)^2} +  \frac{d^{(n-1)}_{\ell- \scriptstyle \frac{1}{2} \displaystyle}}{\ell^2} \right) \sum_{M=-\ell}^{\ell} M^2 \\
    &= \frac{n}{4 d^{(n)}_{\ell}}\left( \frac{\ell }{\ell +1} d^{(n-1)}_{\ell+ \scriptstyle \frac{1}{2} \displaystyle} +  \frac{\ell +1}{\ell} d^{(n-1)}_{\ell - \scriptstyle \frac{1}{2} \displaystyle} \right) \\
    & = \frac{1}{2} + \frac{n}{4} . 
    \end{split}
\end{equation}
For $\ell = 0$ we only have the $\ell = 1/2$ contribution, so that
\begin{equation}
    B_{00} = 0.
\end{equation}
Collecting these results gives
\begin{equation} \label{app:matrixL}
\begin{split} 
    L_{\ell,\ell} &=  \ell (\ell + 1) + \frac{1}{2} \left( \frac{n}{2}+1 \right) x \ \ \forall \ \ell \neq 0, \\
    L_{\ell,\ell+1} &= L_{\ell+1,\ell} = \frac{x}{2} \sqrt{\left( \frac{n}{2} - \ell \right) \left( \frac{n}{2} + \ell + 2 \right)},
\end{split}
\end{equation}
along with $L_{00}=0$ for even $n$.

\begin{figure}
    \centering
    \includegraphics[width=0.48\textwidth]{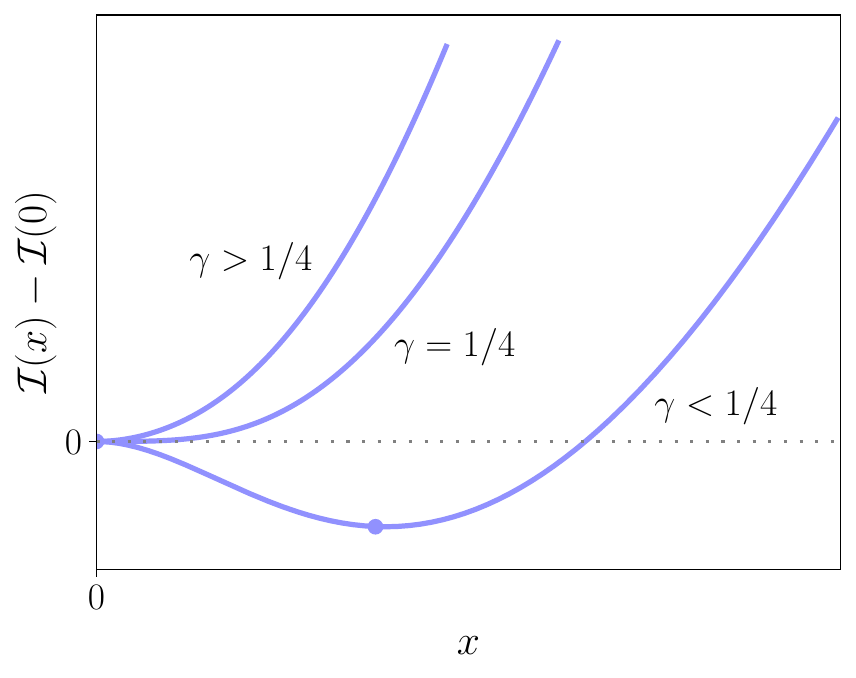} 
    \ \
    \includegraphics[width=0.48\textwidth]{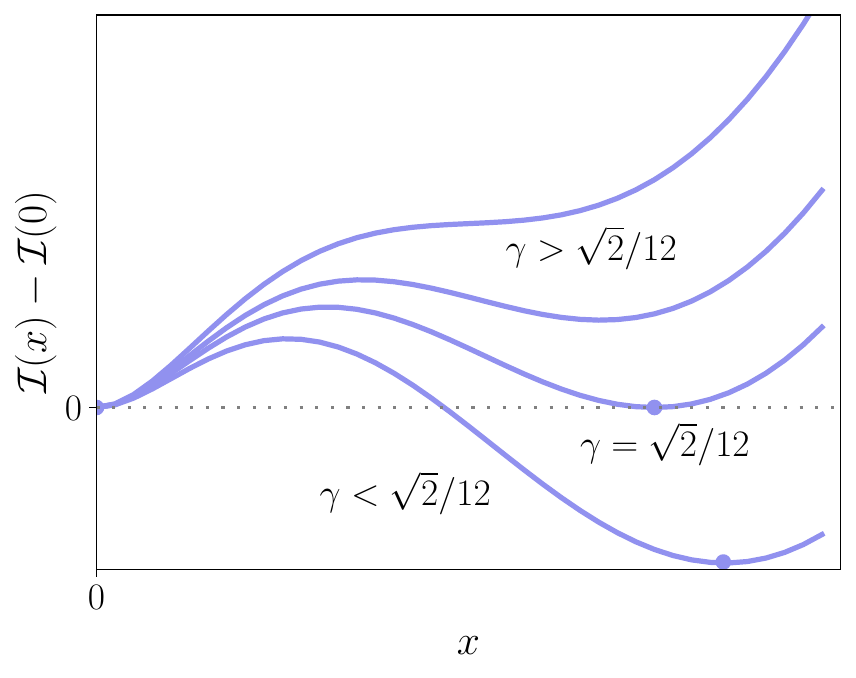}
    \caption{Sketch of the Landau-Ginzburg action $\mathcal{I}(x)$ for $n=2$ (left) and $n=3$ (right). While for $n=2$ we have a second order phase transition, for $n=3$ we have a finite jump.}
    \label{fig:I2I3}
\end{figure}

\subsection{Integer number of replicas
\label{sec:intnumrepl}
}
\subsubsection{\texorpdfstring{Case $n=2$}{Case n=2}}
We have $\ell = 1, 0$ and
\begin{equation}
    L = \begin{pmatrix}
       2 & 0 \\ 0 & 0
    \end{pmatrix} + x \begin{pmatrix}
       1 & \sqrt{3}/2 \\ \sqrt{3}/2 & 0
    \end{pmatrix} . 
\end{equation}
As a consequence 
\begin{equation}
    \Lambda_2 (x) = 1 + \frac{x}{2} + \sqrt{1+x+x^2} , 
\end{equation}
and finally 
\begin{equation}
    \mathcal{I}_2(x) = \frac{2}{1+x} \left[  \left( \gamma + \frac{1}{4} \right) (2+x) - \frac{1}{4} \sqrt{1+x+x^2} \right]^2  . 
\end{equation}
Even if this function is not even in $x$, we have 
\begin{equation}
    \mathcal{I}_2(x) - \mathcal{I}_2(0) =  2 \left( \gamma + \frac{1}{8} \right) \left( \gamma - \frac{1}{4} \right) (x^2-x^3) + O(x^4)
\end{equation}
so that for $\gamma = 1/4$ both the second and the third order vanish. As a consequence we find an Ising-like transition between a broken phase ($\gamma < \gamma_c \equiv 1/4$) and the $x=0$ phase ($\gamma > \gamma_c$) in which the permutational symmetry is restored (see Fig.\,\ref{fig:I2I3}, left panel). This result is consistent with~\cite{bentsen2021measurement}, in which the approximation $n=2$ is implicitly assumed.
\subsubsection{\texorpdfstring{Case $n=3$}{Case n=3}}
We have $\ell = 3/2, 1/2$ and
\begin{equation}
    L = \begin{pmatrix}
       15/4 & 0 \\ 0 & 3/4
    \end{pmatrix} + x \begin{pmatrix}
       5/4 & 1 \\ 1 & 5/4
    \end{pmatrix} \; .
\end{equation}
As a consequence 
\begin{equation}
    \Lambda_3 (x) = \frac{9}{4} + \frac{5}{4}x + \frac{1}{2} \sqrt{9+4 x^2} , 
\end{equation}
and finally 
\begin{equation}
        \mathcal{I}_3(x) = \frac{12}{3+2x} \left[  \left( \gamma + \frac{1}{12} \right) (x+3) + \frac{1}{8} - \frac{1}{8} \sqrt{1+\frac{4}{9}x^2} \right]^2  . 
\end{equation}
$\mathcal{I}_3 (x)$ has a minimum in $x=0$ for every $\gamma > 1/12 \approx 0.083$, and another one, corresponding to a finite $x$, for any $\gamma < 1/24 (1 + 7/\sqrt{13}) \approx 0.123$. The two minima become degenerate for $\gamma = \gamma_c = \sqrt{2}/12 \approx 0.118$ (corresponding to $x=0$, $x=3/2$),  signaling the presence of a discontinuous transition from $x=3/2$ to $x=0$ at $\gamma = \gamma_c$ (see Fig.~\ref{fig:I2I3}, right panel).
\subsubsection{\texorpdfstring{Case $n>3$}{Case n>3}}
\begin{figure}
    \centering
    \includegraphics[width=0.48\textwidth]{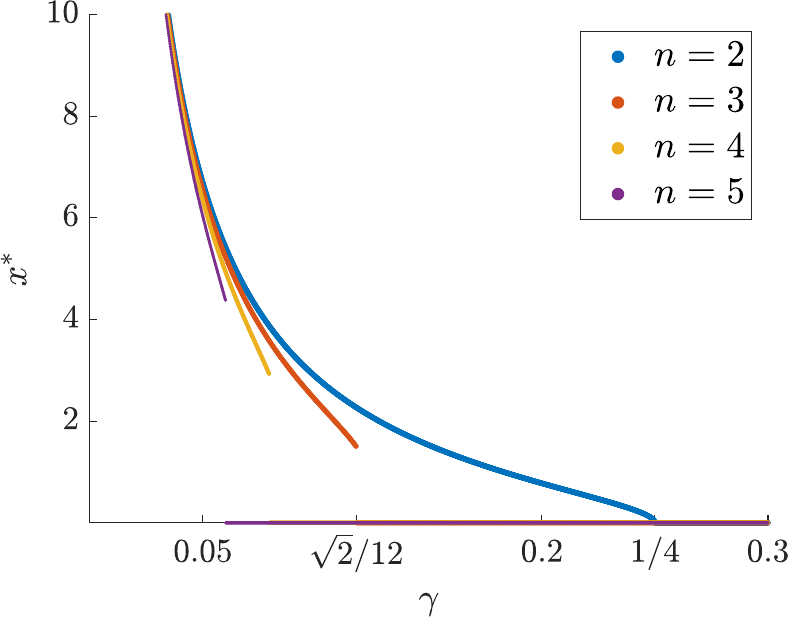} 
    \ \
    \includegraphics[width=0.48\textwidth]{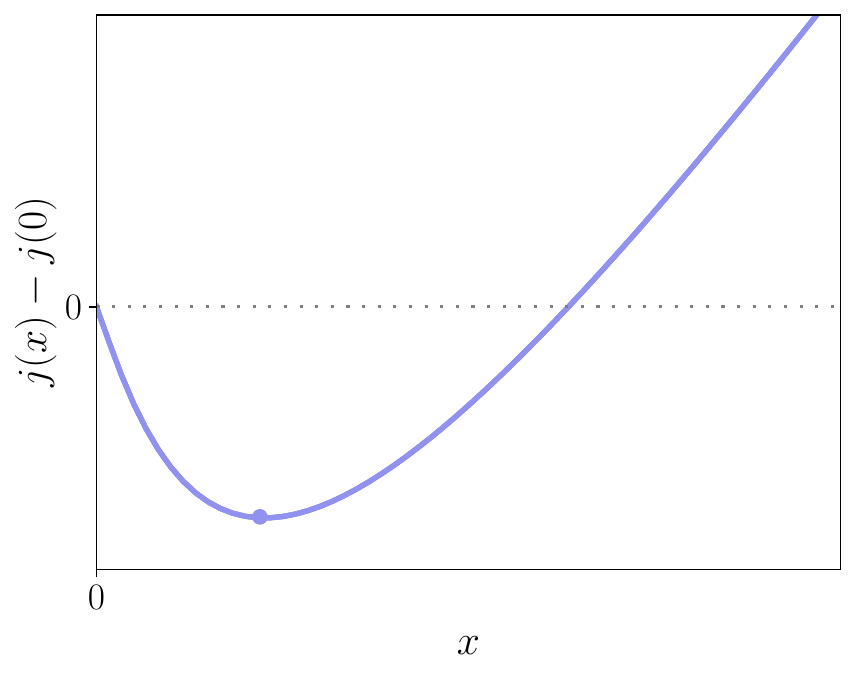}
    \caption{\textit{Left:} Behavior of the minimum $x^{*}$ of the Landau-Ginzburg action $\mathcal{I}(x)$ for $n=2$ to $n=5$. While for any $n$  $x^{*} \equiv 0$ for any $\gamma > \gamma_c$, $x^{*}(\gamma)$ is continuous only for $n=2$. \textit{Right:} Sketch of the Landau-Ginzburg action $\mathcal{I}(x)$ for $n \rightarrow 1^{+}$: regardless of the value of $\gamma$ we have a minimum in correspondence of a finite $x$, signaling that $\gamma_c \rightarrow \infty$ as $n \rightarrow 1^{+}$.}
    \label{fig:I0}
\end{figure}

For $n > 3$ the expression for $\Lambda(x)$ is no longer treatable analytically, so that it is not possible to derive the explicit value of $\gamma_c$. The numerical analysis, however, suggests a scenario similar to the one for $n=3$, so that the phase transition is expected to be discontinuous for any $n \geq 3$ (see Fig.~\ref{fig:I0} left panel).

\subsection{The n-to-1 replica limit}
\makeatletter\protected@edef\@currentlabel{\thesubsection}\makeatother
\label{app:n1limit}
\noindent
As we do not have the generic analytic form of $\Lambda_n(x)$, it is not straightforward to take the $n \rightarrow 1$ limit. As $\Lambda_n(0) = n(n+2)/4$, a perturbative expansion around $x=0$ could be carried out; unfortunately, the higher order ($\geq 3$) coefficients of these expansions turn out to be non-analytic functions of $n$.
In order to obtain an analytical expansion—whether formal or implicit—we will instead consider the expansion for large $x$. At the leading order, $\Lambda_n (x) \sim 3n/4 x$ for $x\gg 1$. Higher orders around $x = \infty$ can be treated with standard methods, obtaining the overall structure
\begin{equation} \label{expansionx}
    \Lambda_n (x) = \frac{3}{4} n (1+x) + n (n-1) \sum^{\infty}_{k=1} P_{k} (n) x^{-k}
\end{equation}
where $P_k(n)$ are increasingly complicated degree $k-1$ polynomials in $n$. In the vicinity of $n=1$ we get
\begin{equation}
    \Lambda_n (x) = \frac{3}{4} n (1+x) + (n-1) \left( \frac{3}{16} \frac{1}{x} - \frac{3}{16} \frac{1}{x^2} + \frac{63}{256} \frac{1}{x^3} - \frac{27}{64} \frac{1}{x^4} + \frac{1899}{2048} \frac{1}{x^5} - \frac{81}{32} \frac{1}{x^6} + \cdots \right) + O(n-1)^2 . 
\end{equation}
This expansion suggests that a non-trivial function of $x$ appears at the first order in $n-1$. In the next subsection, we show how its analytical form can be extracted.
\subsubsection{\texorpdfstring{Direct matrix derivation near $n=1$}{Direct matrix derivation near n=1} \label{sec:besselderiv}}
It is useful to inspect directly the tridiagonal matrix $L$ close to $n=1$. It is convenient to redefine the indices of the matrix introducing
\begin{equation}
    p=\frac{n}{2}-\ell ,
\end{equation}
so that the maximal-spin state corresponds to $p=0$. In this basis the eigenvalue equation
$Lc=\Lambda_n c$ is still tridiagonal. Writing
\begin{equation}
    n=1+\varepsilon,\qquad
    \Lambda_n(x)=\frac34(1+x)+\varepsilon\,\mu(x)+O(\varepsilon^2),
\end{equation}
we can fix $c_0$ as a normalization constant and we find that consistently $c_1 = O(\varepsilon)$, while
\begin{equation}
    c_p=\sqrt{\varepsilon}\,d_p,\qquad p\ge3.
\end{equation}
Using the explicit matrix elements in the $p$ basis, the first three equations of the eigenvalue problem $(L-\Lambda_n)c=0$ read at leading order
\begin{align}
    &(p=0):\qquad \varepsilon\left(1+\frac{x}{4}-\mu(x)\right)c_0+\frac{x}{\sqrt2}c_1=0 \quad \Rightarrow \quad c_1 =\sqrt{2} \varepsilon  c_0 \left(\frac{\mu (x)-1}{x}-1/4\right) \;, \nonumber\\
    &(p=1):\qquad \frac{x}{\sqrt2}(c_0+c_2)-c_1=0, \quad \Rightarrow \quad c_2=-c_0+O(\varepsilon)  \;, \nonumber\\
    &(p=2):\qquad \frac{x}{\sqrt2}c_1+\varepsilon\left(-1+\frac{x}{4}-\mu(x)\right)c_2+\frac{\sqrt3\,x}{2}\varepsilon\,d_3=0 \quad \Rightarrow \quad d_3 = \frac{c_0 (x-4 \mu (x))}{\sqrt{3} x}\;, \\
    &(p=3):\qquad -\sqrt{3} c_0 x+2  \text{i} d_4 x+6 d_3= 0\;. \\
    &(p\geq 4): \qquad     \frac{ \text{i} x}{2}\sqrt{p(p-3)}\,d_{p-1}
    +p(p-2)\,d_p
    +\frac{ \text{i} x}{2}\sqrt{(p-2)(p+1)}\,d_{p+1}=0.
    \label{eq:recpbess0}
\end{align}
The imaginary units have a simple origin: the off-diagonal element $\frac{x}{2}\sqrt{(n/2-\ell)(n/2+\ell+2)}$ has a radicand that turns negative once $n$ is continued to the vicinity of $1$ in the rows $p>3$, so that $\frac{x}{2}\sqrt{p(3-p)}=\frac{\text{i}x}{2}\sqrt{p(p-3)}$ and the analytically continued matrix $L$ is no longer Hermitian.
To solve the recurrence equation for $p\geq 4$, one observes that parameterizing
\begin{equation}
 d_p = D_0(x) e^{\frac{\text{i} \pi  p}{2}} \sqrt{\frac{p-1}{p (p-2)}} K_{p-1} \;,
\end{equation}
the phase $e^{\text{i}\pi p/2}=\text{i}^{\,p}$ exactly cancels the imaginary units, turning Eq.~\eqref{eq:recpbess0} into a three-term recurrence of modified-Bessel type
\begin{equation}
    K_{p+1}=K_{p-1}+\frac{2p}{x}K_p,
\end{equation}
showing that $K_p \equiv K_p(x)$ corresponds with the modified-Bessel function. Finally, the equations for $p=2,3$ give
\begin{equation}
 D_0(x)= -\frac{\text{i} \sqrt{2} c_0}{K_1(x)},\qquad \mu (x)= \frac{1}{4} x \left(\frac{2 K_2(x)}{K_1(x)}+1\right)
 =1+\frac{x}{4}+\frac{x}{2}\frac{K_0(x)}{K_1(x)}
\end{equation}
where in the last step we used the standard modified-Bessel recursion $K_2=K_0+\frac{2}{x}K_1$. Hence
\begin{equation}
    \Lambda_n (x) = \frac{3}{4} (1+x) + (n-1) \left( 1 +  \frac{x}{4} + \frac{x}{2} \frac{K_0 (x)}{K_1(x)}  \right) + O(n-1)^2 .
\end{equation}
Finally, we find 
\begin{equation}
\label{app:IIn1}
    \mathcal{I}_n(x) = \frac{2(n-1)}{1+2x} \left[ \left( \gamma + \frac{1}{4} \right) (1+x) - \frac{x}{4} \frac{K_0(x)}{K_1(x)} \right]^2  + O(n-1)^2 .
\end{equation}
This expression is the first-order expansion of $\mathcal{I}_n(x)$ for $n$ in the proximity of $1$.
As expected the action vanishes for $n=1$, so we inspect the behavior for $n \to 1$. As it happens often in replica limits of disordered systems, the nature of minima/maxima in the Landau-Ginzburg action changes for $n \to 1^+$ and $n \to 1^-$. It is clear that the physical limit for us corresponds to $n \rightarrow 1^{+}$, regardless of the value of $\gamma$, as for $n<1$, the action is not bounded from below. For $n \to 1^+$, the action always has a local maximum for $x=0$ ($\mathcal{I}(x) \sim \mathcal{I}(0) + x^2 \ln x$) and a minimum for finite $x$ (see Fig.\,\ref{fig:I0}, right panel). So, as stated in the main text, the system is always in the broken phase, that is, $x^\ast > 0$ for any $\gamma$. Small and large $x$ expansions of Eq.~\eqref{app:IIn1} give the asymptotic regimes
\begin{align}
&x^\ast \sim \frac{3}{8 \gamma} \to \infty  \;, \qquad \gamma \to 0 \\
&x^\ast \sim 2 e^{-2 \gamma -\gamma_{\rm EM} -1} \;, \qquad \gamma \to \infty
\end{align}
where $\gamma_{\rm EM}$ is the Euler-Mascheroni constant. 
\subsection{The n-to-0 replica limit}
\noindent
To compute the $n \rightarrow 0$ limit, we can start again from Eq.\,\eqref{expansionx}. Continuing the coefficients $P_k(n)$ to $n=0$, we get
\begin{equation}
    \Lambda_n (x) = n \left( \frac{3}{4}(1+x) - \frac{3}{16} \frac{1}{x} + \frac{9}{32} \frac{1}{x^2} - \frac{153}{256} \frac{1}{x^3} + \frac{819}{512} \frac{1}{x^4} - \frac{10449}{2048} \frac{1}{x^5} + \frac{76977}{4096} \frac{1}{x^6} \dots \right) + O(n^2)
\end{equation}
in which we can recognize the asymptotic expansion of
\begin{equation}
    \Lambda_n (x) = \frac{n}{2} \left( \frac{K_1(x)}{K_1(x)-K_0 (x)} - \frac{x}{2}   \right) + O(n^2) \ .
\end{equation}
Injecting this result in the action, we have:
\begin{equation}
    \mathcal{I}(x) = - \frac{n}{x} \left[ \left( \gamma - \frac{1}{2} \right) x + \frac{1}{4} \frac{K_1(x)}{K_1(x)-K_0(x)} \right]^2  + O(n^2) \ .
\end{equation}
In this case the only physical choice is to  send $n \rightarrow 0^{-}$ as, in the other case, $\mathcal{I}(x)$ is not bounded from below. Once again, we get the broken solution $x = x^{*}(\gamma) > 0$ for every value of $\gamma$, so that no phase transition occurs in the forced-measurement ensemble either. The asymptotic behaviors parallel those of the $n \to 1$ limit: for $\gamma \to 0$ the minimum is found at large $x$, with $x^\ast \simeq 3/(8\gamma)$, while for $\gamma \gg 1$ it moves to small $x$, with $x^\ast \simeq 1/(4\gamma)$ up to logarithmic corrections.

\newpage 

\section{Decoupling of replicas}

\subsection{Derivation of the single-spin unraveling}
\makeatletter\protected@edef\@currentlabel{\thesubsection}\makeatother
\label{sec:replicalimitunraveling}
\noindent
In this subsection we derive Eq.~\eqref{main:drho2} and the corresponding
self-consistency conditions in a form adapted to the $n\to1$ limit. We start
from the reduced single-site problem obtained after inserting the Ansatz
\eqref{app:ansatzX} into the replicated path integral. The resulting boundary
matrix element can be written as
\begin{equation} 
    \Tr [\Omega^{(n)} \trho1^{(n)}(T)] = \int \mathcal{D}[\mathbf{s}] e^{\mathcal{S}^{(1)}} \ \bra{\mathbf{s}^{-}_{a} (T)} \Omega^{(n)} \ket{\mathbf{s}^{+}_{a} (T)} \bra{\mathbf{s}^{+}_{a} (0)} \trho1^{(n)}_0 \ket{\mathbf{s}^{-}_{a} (0)} . 
\end{equation}
In the replica-symmetric sector the single-site action becomes
\begin{equation}
\begin{split}
    \mathcal{S}^{(1)} &=
    \sum_{a, \sigma} \text{i} \sigma \mathcal{S}_{\rm top}[\mathbf{s}_{a,\sigma}]
    +  \int dt \ \frac{2\gamma + r - q}{4}  \left( \sum_{a, \sigma} \mathbf{s}_{a \sigma} \right)^2
    -  \int dt \ \frac{r + q}{4} \left( \sum_{a,\sigma} \sigma \mathbf{s}_{a \sigma} \right)^2 \\
    &\quad
    + \sum_{a} \int dt \ X \  \mathbf{s}_{a +} \cdot \mathbf{s}_{a-}
    + \frac{3}{4} n \int dt \left( q -  \frac{3}{4} - 2 \gamma \right)   .
\end{split}
\end{equation}
Applying a second Hubbard-Stratonovich decoupling to the two collective
quadratic terms,
\begin{equation}
        e^{A^2/4} \sim \int dh e^{-h^2 + h A} \hspace{1cm} e^{-A^2/4} = \int dh e^{-h^2 + ihA}
\end{equation}
one decouples the replicas and obtains
\begin{equation} \label{app:h1h2}
    \Tr [\Omega^{(n)} \trho1^{(n)}(T)] = \int \mathcal{D}[\mathbf{h}]  e^{- \frac{1}{2} \int dt \left[\mathbf{h_1}^2/\gamma_1 + \mathbf{h_2}^2/\gamma_2 \right]} \int  \mathcal{D}[\mathbf{s}] e^{ \sum_a \tilde{\mathcal{S}} [\mathbf{s}_{a,\sigma}]} \ \bra{\mathbf{s}^{\sigma}_{a} (T)} \Omega^{(n)} \ket{\mathbf{s}^{\sigma}_{a} (T)} \bra{\mathbf{s}^{+}_{a} (0)} \trho1^{(n)}_0 \ket{\mathbf{s}^{-}_{a} (0)} . 
\end{equation}
where
\begin{equation}
    \gamma_1 = \gamma + \frac{1}{2} (r - q) \hspace{1cm} \gamma_2  = \frac{1}{2} (r + q) \ 
\end{equation}
and
\begin{equation}
\label{eq:singlereplicapathint}
    \tilde{S}[\mathbf{s}_{a,\sigma}] = \sum_{\sigma} \int dt \left( \text{i} \sigma \mathcal{S}_{\rm top}[\mathbf{s}_{a,\sigma}] + (\mathbf{h}_1 + \text{i} \sigma \mathbf{h}_2) \cdot \mathbf{s}_{a,\sigma} \right) + X \int dt \ \mathbf{s}_{a,+} \cdot \mathbf{s}_{a,-}  + \frac{3}{4} \left( q -  \frac{3}{4} - 2 \gamma \right) T . 
\end{equation}
Equation~\eqref{eq:singlereplicapathint} is the path-integral representation of
the time evolution of the unnormalised single-spin density matrix $\trho1$,
\begin{equation}
\label{app:tdrho1}
    \trho1 + d\trho1 = e^{- \text{i} d\mathscr{H}} \trho1
    e^{ \text{i} d\mathscr{H}^\dag}  + X \sum_\alpha \mathcal{D}_{S^\alpha}[\trho1] \;, \qquad d\mathscr{H} = (\sqrt{\gamma_2} dW^\alpha + \text{i} \sqrt{\gamma_1} dM^\alpha)  \hat S^\alpha
\end{equation}
where additive constants have been ignored and $\sqrt{\gamma_1} dM^\alpha \sim
h_1^\alpha dt$, $\sqrt{\gamma_2} dW^\alpha \sim h_2^\alpha dt$. Here the Hermitian
part $\sqrt{\gamma_2}\,dW^\alpha \hat S^\alpha$ generates the unitary (commutator)
noise, while the anti-Hermitian part $i\sqrt{\gamma_1}\,dM^\alpha \hat S^\alpha$
generates the measurement (anticommutator) backaction; consistently,
$\gamma_1=\gamma+(r-q)/2$ carries the bare monitoring rate $\gamma$. In the replica
limit $n \to 1$, this becomes the effective single-spin monitored dynamics,
\begin{equation} 
\label{app:drho1}
    d \varrho = \left( \frac{3}{4} + \gamma \right) \left( \hat S^{\alpha} \varrho \hat S^{\alpha} - \frac{3}{4} \varrho \right) +  \sqrt{\gamma_1} dY^{\alpha} \lbrace \hat S^{\alpha} - \braket{\hat S^{\alpha}}, \varrho \rbrace - \text{i} \sqrt{\gamma_2} dW^{\alpha} [\hat S^{\alpha},\varrho]  \ .
\end{equation}
Passing from $\trho1$ to the normalized state $\varrho=\trho1/\tr\trho1$ leaves the
unitary noise $W$ untouched and only reweights the measurement channel: the unbiased
increment $M$ is replaced by the physical innovation $Y$, the Born subtraction
$\hat S^{\alpha}\to\delta\hat S^{\alpha}=\hat S^{\alpha}-\braket{\hat S^{\alpha}}$
enforcing $\tr\varrho=1$ at all times.
Introducing the corresponding noise-dependent propagators,
\begin{equation}
\label{eq:twopropagators}
    \trho1_{t_2} = \tilde{\mathcal{U}}_{t_1, t_2}^{(\mathbf{W}, \mathbf{M})} [\trho1_{t_1}]\;, \qquad
    \varrho_{t_2} = \mathcal{U}_{t_1, t_2}^{(\mathbf{W}, \mathbf{Y})} [\varrho_{t_1}] \;.
\end{equation}
The two maps are of a very different nature. For a fixed realization of the noises,
$\tilde{\mathcal{U}}_{t_1, t_2}^{(\mathbf{W}, \mathbf{M})}$ is the propagator of the
\emph{unnormalized} state \eqref{app:tdrho1}: since that equation contains no Born
subtraction, it acts \emph{linearly} on $\trho1$, and it is this linearity that
allows the composition $\tilde{\mathcal{U}}_{[t,T]}\tilde{\mathcal{U}}_{[0,t]}$ and the
factorization used above. By contrast,
$\mathcal{U}_{t_1, t_2}^{(\mathbf{W}, \mathbf{Y})}$ propagates the \emph{normalized}
state through the SSE \eqref{app:drho1}, which is \emph{nonlinear}: it depends on the
running expectation value $\braket{\hat S^\alpha}=\tr(\varrho\,\hat S^\alpha)$ via
$\delta\hat S^\alpha$, and the physical innovation $Y$ is itself defined along the
trajectory. Accordingly $\mathcal{U}^{(\mathbf{W},\mathbf{Y})}[\,\cdot\,]$ is not a
linear operator on its argument and is meaningful only along a given physical
trajectory. This is why the algebraic manipulations (splitting at the intermediate
time $t$, factorization of the overlap) are performed at the level of the linear map
$\tilde{\mathcal{U}}^{(\mathbf{W},\mathbf{M})}$, the passage to the nonlinear physical
dynamics being made only at the end through the Born reweighting.
One then has for any $k$-replica operator $\Omega_k$
\begin{equation}
\label{app:replicaeq}
    E_{\mathbf{W}, \mathbf{M}}( \Tr(\Omega_k \trho1(t)^{\otimes k}) \tr(\trho1(t))^{1-k}) =
    E_{\mathbf{W}, \mathbf{Y}}( \Tr(\Omega_k \varrho(t)^{\otimes k})) 
\end{equation}
where we used $X + r = 3/4$, required for trace preservation. The Wiener processes
$M^\alpha$ and $W^\alpha$ are effective single-site noises emerging at the
mean-field level and are not the original many-body noises of Eq.~\eqref{eq:ham}.
Via $\gamma_1,\gamma_2$, the evolution \eqref{app:drho1} still depends on the
time-dependent saddle fields $q,r,X$.
To proceed further, we focus on the case where $\Omega^{(n)} = \mathbb{1}$, which is relevant for the calculation of the moments of few-body observables. Using that $X_{ab}^{++} = q$ and $X_{ab}^{+-} = r$ for $a \neq b$, we have
\begin{equation} \label{eq:saddlepointgeneral}
    \begin{split}
         &q (t) = E_{\mathbf{W}, \mathbf{M}} \left(\sum_\alpha\frac{   \tr  \left( \ \tilde{\mathcal{U}}_{[t, T]} \bigl[ \hat S^\alpha\tilde{\mathcal{U}}_{[0,t]}[\trho1_0] \bigr]\right)  \tr  \left( \ \tilde{\mathcal{U}}_{[t, T]} \bigl[ \hat S^\alpha\tilde{\mathcal{U}}_{[0,t]}[\trho1_0] \bigr]\right) }{\tr\left( \ \tilde{\mathcal{U}}_{[0, T]} [\trho1_0]\right) } \right) \\
   &r(t) = \frac{3}{4} - X(t) =  E_{\mathbf{W}, \mathbf{M}} \left(\sum_\alpha\frac{   \tr  \left( \ \tilde{\mathcal{U}}_{[t, T]} \bigl[ \hat S^\alpha\tilde{\mathcal{U}}_{[0,t]}[\trho1_0] \bigr]\right)  \tr  \left( \ \tilde{\mathcal{U}}_{[t, T]} \bigl[ \tilde{\mathcal{U}}_{[0,t]}[\trho1_0] \bigr] \hat S^\alpha\right) }{\tr\left( \ \tilde{\mathcal{U}}_{[0, T]} [\trho1_0]\right) } \right) \ 
    \end{split}
\end{equation}
To write down explicitly the saddle-point equations directly in terms of \eqref{app:drho1}, we could make use of Eq.~\eqref{app:replicaeq}, however the insertion of the operators $S^\alpha_a$ at a time $t \in [0, T]$ requires a slightly different approach. We use that the evolutions $\tilde{\mathcal{U}}_{[0,t]}$ and $\tilde{\mathcal{U}}_{[t, T]}$ are statistically independent --- the effective single-site noises are white, so the two disjoint time intervals share no increments --- and that the time-reversal symmetry of the bulk generator (the relation $\hat{\mathbb U}^{(n)\dagger}_{X;t_2,t_1}=\hat{\mathbb U}^{(n)}_{\mathcal T X;t_2,t_1}$ with $\mathcal T X(t)=X(T-t)^\ast$ derived above) implies the reflection property $q(t) = q(T-t)$, $r(t) = r(T-t)$, $X(t) = X(T-t)$. We denote
\begin{equation}
    \trho1_t^{(1)} = \tilde{\mathcal{U}}_{[0,t]}(\trho1_0), \qquad
    2 \trho1_{T-t}^{(2)} = \tilde{\mathcal{U}}_{[0,T-t]}(\mathbb{1}).
\end{equation}
The second definition is chosen so that, for any operator $A$ inserted at time $t$,
\begin{equation}
\label{app:timereversalAux}
    \tr\!\left(\tilde{\mathcal{U}}_{[t,T]}[A]\right)=2\tr\!\left(\trho1^{(2)}_{T-t} A\right).
\end{equation}
This identity is precisely where time reversal enters explicitly. Writing
$\tr(\tilde{\mathcal{U}}_{[t,T]}[A])=\tr(A\,\tilde{\mathcal{U}}^{\dagger}_{[t,T]}[\mathbb{1}])$
in terms of the adjoint (Heisenberg) propagator, the factor
$\tilde{\mathcal{U}}^{\dagger}_{[t,T]}[\mathbb{1}]$ is the backward evolution of
the identity from $T$ down to $t$. Through the time-reversal symmetry of the bulk
generator recalled above and the reflection $t\to T-t$ of the saddle fields, this
backward evolution has the same law as the \emph{forward} evolution
$\tilde{\mathcal{U}}_{[0,T-t]}[\mathbb{1}]=2\,\trho1^{(2)}_{T-t}$ of the maximally
mixed state. In other words, the segment $[t,T]$ read in reverse is statistically
equivalent to an independent copy of the dynamics run forward for a time $T-t$,
which is what allows us to treat $\trho1^{(1)}_{t}$ and $\trho1^{(2)}_{T-t}$ as
two independent realizations of the \emph{same} stochastic evolution. Therefore
\begin{equation}
    \tr\!\left( \tilde{\mathcal{U}}_{[0, T]} [\trho1_0]\right)
    = \tr\!\left( \tilde{\mathcal{U}}_{[t, T]}\tilde{\mathcal{U}}_{[0, t]} [\trho1_0]\right)
    = 2 \tr\!\left(\trho1_{T-t}^{(2)} \trho1_{t}^{(1)}\right).
\end{equation}
The quantity $\tr\!\left( \tilde{\mathcal{U}}_{[0, T]} [\trho1_0]\right)$ is precisely the
Born weight of a trajectory in the Gaussian measure $E_{\mathbf{W},\mathbf{M}}$.
The factorization above shows that, after splitting the evolution at the intermediate
time $t$, this weight can be written as the overlap of the two independent
unnormalized states $\trho1_t^{(1)}$ and $\trho1_{T-t}^{(2)}$. Using this identity, we can rewrite
\begin{equation}
\label{app:selfconstq}
    q(t) = 
    E_{\mathbf{W}, \mathbf{M}} \left(\frac{  2 \sum_\alpha \tr  \left(\trho1_{T-t}^{(2)} \hat S^\alpha \trho1_{t}^{(1)}\right)^2}{\tr(\trho1_{T-t}^{(2)} \trho1_{t}^{(1)})} \right).
\end{equation}
Writing $\trho1^{(k)}=z_k\varrho^{(k)}$ with $z_k=\tr(\trho1^{(k)})$, the overlap in the
denominator becomes $\tr(\trho1_{T-t}^{(2)} \trho1_t^{(1)})=z_1 z_2 \tr(\varrho_{T-t}^{(2)}\varrho_t^{(1)})$,
while each trace in the numerator carries one factor $z_1$ and one factor $z_2$; the ratio in
Eq.~\eqref{app:selfconstq} therefore retains a single overall factor $z_1 z_2$ relative to its
normalized counterpart. Each $z_k=\tr\trho1^{(k)}$ is precisely the Born weight of the
corresponding trajectory segment, cf.~Eq.~\eqref{eq:Frhoreplica}. Reweighting the unbiased
Gaussian measure by these Born factors is exactly the operation \eqref{app:replicaeq} that turns
the Gaussian average $E_{\mathbf{W},\mathbf{M}}$ into the physical average
$E_{\mathbf{W},\mathbf{Y}}$; since the two segments are independent, the reweighting factorizes
and acts on each of them separately, $E_{\mathbf{W},\mathbf{M}}(z_1 z_2\,\cdots)=E_{\mathbf{W},\mathbf{Y}}(\cdots)$.
We therefore obtain
\begin{equation}
    q(t)=2 E_{\mathbf{W}, \mathbf{Y}} \left(\frac{\sum_\alpha\tr[\varrho^{(2)}_{T-t} \hat S^\alpha \varrho^{(1)}_t]^2}{\tr[\varrho^{(2)}_{T-t} \varrho^{(1)}_t]}\right).
\end{equation}
In the following, for brevity, we denote this physical average simply by $E(\cdots)$. The stochastic variables $\varrho^{(1)}(t)$ and $\varrho^{(2)}(T-t)$ correspond to two independent evolutions of the initial single-site density matrix $\varrho = \mathbb{1}/2$ with Eq.~\eqref{app:drho1} up to times $t$ and $T-t$, respectively. Similarly, for 
\begin{align} 
&\frac{3}{4}  - X(t) = 2  E \left( \frac{\sum_\alpha\tr[\varrho^{(2)}_{T-t} \hat S^\alpha \varrho^{(1)}_t]\tr[\varrho^{(1)}_t \hat S^\alpha \varrho^{(2)}_{T-t}]}{\tr[\varrho^{(2)}_{T-t}\varrho^{(1)}_t]}\right) \ .
\end{align}

\subsection{Derivation of the stationary distribution}
\makeatletter\protected@edef\@currentlabel{\thesubsection}\makeatother
\label{app:singleunravel}
\noindent
In this subsection we derive Eq.~\eqref{main:Pstat}. In the limit $T \to \infty$ and for $0\ll t \ll T$, the values of the parameters $X(t)$ and $q(t)$ reach a plateau, and both $\varrho^{(1)}(t)$ and $\varrho^{(2)}(T-t)$ are drawn from the same stationary distribution. We now derive the explicit form of this stationary distribution, given in \eqref{main:Pstat} in the main text. Let us parameterise the state of the single spin as
\begin{equation}
\label{eq:singlespinrho}
    \varrho = \frac{\ide}{2} +  \mathsf{r}_\alpha \hat S^\alpha \;, \qquad \tr ( \varrho \hat S^\alpha ) = \frac{1}{2} \mathsf{r}_\alpha  \ , 
\end{equation}
in terms of a vector $\mathsf{r}^\alpha$. Positivity requires $\mathsf{r} = \sqrt{\mathsf{r}_\alpha \mathsf{r}_\alpha} = \sqrt{R} \leq 1$ and the state is pure for $\mathsf{r} = 1$. Plugging this into Eq.~\eqref{app:drho1}, we obtain stochastic equations for $\mathsf{r}_\alpha$
\begin{equation}
\label{eq:nalpha}
    d \mathsf{r}_\alpha = 2 \tr[d \varrho \hat S^\alpha]=  \sqrt{\gamma_2} \epsilon_{\alpha \beta \gamma} dW^\beta \mathsf{r}_\gamma  + \sqrt{\gamma_1} dY^\alpha -\sqrt{\gamma_1} \mathsf{r}_\alpha 
    dY^\beta \mathsf{r}_\beta - (\gamma + 3/4) \mathsf{r}_\alpha dt \;.
\end{equation}
Analysing this equation, it is clear that it describes the motion of a vector $\mathsf{r}_\alpha$, whose angular part diffuses isotropically.
Instead, for its squared length $R = \mathsf{r}_\alpha \mathsf{r}_\alpha$, we obtain after some manipulations
\begin{equation}
    dR = 2 \sqrt{\gamma_1} (R^{1/2} - R^{3/2}) dB + \gamma_1 \left(R^2-4 R+3\right) dt -2 X R  dt \ . 
\end{equation}
where we introduced $dY^\alpha \mathsf{r}_\alpha = \sqrt{\mathsf{r}_\alpha \mathsf{r}_\alpha} dB = \sqrt{R} dB$, with $dB$ a single Wiener Process with $dB^2 = dt$ (Wiener processes can be summed in quadrature).
Finally, in terms of $\mathsf{r} = \sqrt{R}$
\begin{equation}
    d\mathsf{r} = dt \left(\frac{\gamma_1}{\mathsf{r}}-\mathsf{r} (\gamma_1+X)\right)+\sqrt{\gamma_1} dB \left(1-\mathsf{r}^2\right)
\end{equation}
To find the stationary distribution, it is useful to make a change of variable and recast this equation in the Langevin form. We set 
\begin{equation}
    [0, 
    \infty) \ni \omega = \arctanh(\mathsf{r}) \;, \qquad \mathsf{r} = \tanh(\omega) \in [0,1) \ , 
\end{equation}
so that the new variable $\omega$ satisfies the Langevin equation
\begin{equation}
    d\omega = \sqrt{\gamma_1} dB - dt V'(\omega)
\end{equation}
with the potential
\begin{equation}
    V(\omega) = \frac{1}{2} X \cosh ^2(\omega)-\gamma_1\log (\sinh (2 \omega)) \ . 
\end{equation}
We can thus directly write down the stationary distribution
\begin{equation}
    P_{\rm stat}(\omega) = \frac{1}{Z} e^{- 2V(\omega) / \gamma_1} = \frac{1}{Z} \sinh ^2(2 \omega) e^{- 2x \cosh ^2(\omega)} \ , 
\end{equation}
where we used the fact that that $x = X/(2 \gamma_1)$. Equivalently, in terms of the variable $\mathsf{r}$
\begin{equation}
\label{app:steadyP}
    P_{\rm stat}(\mathsf{r}) = P_{\rm stat}(\omega) \frac{d\omega}{d\mathsf{r}} = \frac 1 Z \frac{ \mathsf{r}^2}{\left(1-\mathsf{r}^2\right)^3}  e^{- 2x(1-\mathsf{r}^2)^{-1}}
\end{equation}
Note that as expected, if $x \to 0$, the distribution becomes more and more peaked around $\mathsf{r} = 1$, implying that the stationary distribution is a random pure state on the Bloch sphere.  

\subsection{Self-consistent equation for x}
\makeatletter\protected@edef\@currentlabel{\thesubsection}\makeatother
\label{par:selfconst}
\noindent
We now show that the self-consistent equations for $x$ in the bulk, obtained from Eq.~(15), coincide with the condition obtained by minimizing the LG action \eqref{main:mathIx}.

Plugging the result of Eq.\,\eqref{main:Pstat}, we find the self-consistent equations for $q$ and $X$ in the bulk. Parameterising $\varrho^{(k)} = 1/2 + \mathsf{r}_\alpha^{(k)} S^\alpha$ and denoting by $\theta$ the angle between the vectors $\mathsf{r}^{(1)}_\alpha$ and $\mathsf{r}^{(2)}_\alpha$, we can express
\begin{equation}
\begin{split}
\tr[\varrho^{(1)} \varrho^{(2)}] &= \frac{1}{2}( 1 + \mathsf{r}_1 \mathsf{r}_2 \cos(\theta)) \;, \\
\sum_\alpha\tr[\varrho^{(1)} \hat S^\alpha \varrho^{(2)}]^2 &= \frac{1}{16} [ \mathsf{r}_1^2 + \mathsf{r}_2^2 + 2 \mathsf{r}_1 \mathsf{r}_2 \cos(\theta)  - \sin(\theta)^2 \mathsf{r}_1^2 \mathsf{r}_2^2
]  \\
\sum_\alpha\tr[\varrho^{(1)} \hat S^\alpha \varrho^{(2)}]\tr[\varrho^{(2)} \hat S^\alpha \varrho^{(1)}] &= \frac{1}{16} [ \mathsf{r}_1^2 + \mathsf{r}_2^2 + 2 \mathsf{r}_1 \mathsf{r}_2 \cos(\theta)  + \sin(\theta)^2 \mathsf{r}_1^2 \mathsf{r}_2^2
] \ .
\end{split}
\end{equation}
Averaging over $\theta$, we are left with 
\begin{align}
    q &= \frac1 4 -\int d\mathsf{r}_1 d\mathsf{r}_2 P_{\rm stat}(\mathsf{r}_1) P_{\rm stat}(\mathsf{r}_2) \frac{\left(1-\mathsf{r}_1^2\right) \left(1-\mathsf{r}_2^2\right) \arctanh(\mathsf{r}_1 \mathsf{r}_2)}{4 \mathsf{r}_1 \mathsf{r}_2} \\ 
    \frac{3}{4} - X &= \frac3 4 -\int d\mathsf{r}_1 d\mathsf{r}_2 P_{\rm stat}(\mathsf{r}_1) P_{\rm stat}(\mathsf{r}_2) \frac{\left(3 -\mathsf{r}_1^2 \mathsf{r}_2^2-\mathsf{r}_1^2-\mathsf{r}_2^2\right) \arctanh(\mathsf{r}_1 \mathsf{r}_2)}{4 \mathsf{r}_1 \mathsf{r}_2}
\end{align}
or equivalently
\begin{equation}
\begin{split}
    2 \gamma_1 &= 2 \gamma +  \frac 1 2 + \frac 1 2 \int d\mathsf{r}_1 d\mathsf{r}_2 P_{\rm stat}(\mathsf{r}_1) P_{\rm stat}(\mathsf{r}_2) \left(\mathsf{r}_1 \mathsf{r}_2 - \frac{1}{\mathsf{r}_1 \mathsf{r}_2}\right)\arctanh(\mathsf{r}_1 \mathsf{r}_2) \;, \\
    X &= \int d\mathsf{r}_1 d\mathsf{r}_2 P_{\rm stat}(\mathsf{r}_1) P_{\rm stat}(\mathsf{r}_2) \frac{\left(3 -\mathsf{r}_1^2 \mathsf{r}_2^2-\mathsf{r}_1^2-\mathsf{r}_2^2\right) \arctanh(\mathsf{r}_1 \mathsf{r}_2)}{4 \mathsf{r}_1 \mathsf{r}_2}
\end{split}
\end{equation}
Those integrals can be evaluated by means of the change of variable
$\mathsf{r} = \sqrt{u /(1+u)}$, which leads to
\begin{equation}
    P_{\rm stat} (u) = \frac{1}{Z} \sqrt{u(1+u)} e^{-x(2u+1)} \;, \quad Z = \frac{K_1(x)}{4 x}
\end{equation}
In terms of the new variable we have 
\begin{equation}
\begin{split}
  2 \gamma_1 &= 2 \gamma +  \frac 1 2 + \frac{1}{4Z^2} \partial_x \ \int_0^{\infty} du_1 du_2 e^{-2x(u_1+u_2+1)} \arctanh \sqrt{\frac{u_1 u_2}{(1+u_1)(1+u_2)}} \\
    X &= \frac{1}{4 Z^2} \left( 1 - \partial_x \right)  \int_0^{\infty} du_1 du_2 e^{-2x(u_1+u_2+1)} \arctanh \sqrt{\frac{u_1 u_2}{(1+u_1)(1+u_2)}} . 
\end{split}
\end{equation}
This last integral appearing in both equations can be computed integrating by parts in the variable $u_1$
\begin{equation}
\begin{split}
    &\int_0^{\infty} du_1 du_2 e^{-2x(u_1+u_2+1)} \arctanh \sqrt{\frac{u_1 u_2}{(1+u_1)(1+u_2)}} \\ 
    &= \frac{1}{4x} \int_0^{\infty} du_1 du_2 \frac{ e^{-2x(u_1+u_2+1)}}{u_1 + u_2 + 1} \sqrt{ \frac{u_2 (1+u_2)}{u_1 (1+u_1)}} \\
    &= \frac{1}{2x} \int_x^{\infty} ds \int_0^{\infty} du_1 du_2 e^{-2s(u_1+u_2+1)}  \sqrt{ \frac{u_2 (1+u_2)}{u_1 (1+u_1)}} \\
    &= \frac{1}{8x} \int_x^{\infty} ds \frac{K_1(s) K_0(s)}{s} \\
    &= -\frac{1}{8x} K_0(x) K_1(x) + \frac{1}{8} ( K_0(x) K_2(x) - K_1(x)^2) , 
\end{split}
\end{equation}
from which, setting $k(x) = \frac{K_0(x)}{K_1(x)}$, we obtain
\begin{equation}
    \begin{split}
         2 \gamma_1 &= 2 \gamma + \frac{1}{2} \left[1 - x ( k(x)^2 - 1 ) \right] - k(x)\\
        X &= \frac{1}{2} \left[2 k(x) + x k(x) + x(1+x) ( k(x)^2 - 1 ) \right] \ . 
    \end{split}
\end{equation}
Taking the ratio of these two equations and using $x = X/(2 \gamma_1)$, we get 
\begin{equation}
    (2+3x) k(x) + x (1+2x)( k(x)^2 - 1 ) = (1+ 4 \gamma) x . 
\end{equation}
Using standard properties of Bessel functions, we have $x k'(x) = k(x) + x ( k(x)^2 - 1 )$, so that we find
\begin{equation} \label{app:selfconstx}
    (1+x) k(x) + x(1+2x) k^{\prime}(x) = (1+ 4 \gamma)x \ . 
\end{equation}
This condition is exactly equivalent to $\mathcal{I}^{\prime} (x) = 0$, with $\mathcal{I}(x)$ given by Eq.\,\eqref{main:In=1}. Indeed, differentiating Eq.\,\eqref{main:In=1} and using the identity above, one finds
\begin{equation}
    \mathcal{I}'(x) = - \frac{\left[(1+4\gamma)(1+x)-x k(x)\right]\left[(2+3x)k(x)+x(1+2x)\left(k(x)^2-1\right)-(1+4\gamma)x\right]}{(1+2x)^2} \; ,
\end{equation}
and the second factor is precisely Eq.\,\eqref{app:selfconstx}. Therefore the solution obtained from the analytic continuation of the replica action coincides analytically with the one obtained from the single-spin unraveling.

\subsection{Derivation of the purity bound}
\noindent 
In this subsection we derive Eq.~\eqref{main:purity}, namely the lower bound for the purity $\Pi$. To calculate $\Pi$, we have to replace $\riesz^{(n)}$ with $\Swap \otimes \ide^{\otimes(n-2)}$, with $\Swap$ the swap operator of two replicas. Thus, the boundary conditions belong to different symmetry sectors (the swap and identity permutation), and in a broken phase the saddle-point solution must exhibit at least one instanton. The exact value of the instanton action requires solving the saddle-point equations with the specific boundary conditions and corresponds to the minimum of the action. However, an upper bound can be obtained by imagining that the instanton occurs instantaneously at an arbitrary time point $0 \ll t \ll T$: in this case, the corresponding value of the action can be obtained from the overlap between the steady state in the identity sector and the steady state in the swap sector. We recall that the steady states for different sectors are obtained via Eq.~\eqref{eq:othersectors}. By unraveling this quantity
as explained in Sec.~\ref{sec:replicalimitunraveling}, in terms of the solutions $\varrho$ of \eqref{app:drho1}, we obtain $\Pi \sim T e^{-N \mathcal{I}^{*}}$. In the resulting overlap formula, the numerator compares the stationary states in the identity and swap sectors, while the denominator is the normalization inherited from the replica unraveling:
\begin{equation}
     e^{-\mathcal{I}^{*}} \gtrsim 2 E \left(   \frac{\Tr^{(2)}{[(\varrho^{(1)}\otimes\varrho^{(2)}) \mathbb{S} (\varrho^{(1)}\otimes\varrho^{(2)})]}}{\tr(\varrho^{(1)} \varrho^{(2)})} \right)_{\varrho_1,\varrho_2} =
     2 E \left(   \frac{\tr{(\varrho^{(1)}\varrho^{(2)}\varrho^{(1)}\varrho^{(2)})}}{\tr(\varrho^{(1)} \varrho^{(2)})} \right)_{\varrho_1,\varrho_2} 
\end{equation}
where we are averaging over $\varrho^{(1)}, \varrho^{(2)}$ drawn independently from the stationary distribution \eqref{app:steadyP}. 
Parameterising $\varrho^{(1)}$ and $\varrho^{(2)}$ as in Eq.~\eqref{eq:singlespinrho} in terms of two vectors $\boldsymbol{\mathsf{r}}^{(1)}, \boldsymbol{\mathsf{r}}^{(2)}$, we get
\begin{align}
    &\tr[\varrho^{(1)} \varrho^{(2)}]= \frac{1}{2} \left( 1 + \mathsf{r}^{(1)} \mathsf{r}^{(2)} \cos(\theta) \right) \;,\\
    &\tr(\varrho^{(1)} \varrho^{(2)}\varrho^{(1)} \varrho^{(2)})  = \frac{1}{8} \left((\mathsf{r}^{(1)})^2
    + (\mathsf{r}^{(2)})^2+1+
    \mathsf{r}^{(1)} \mathsf{r}^{(2)} (4 \cos (\theta )+\mathsf{r}^{(1)} \mathsf{r}^{(2)} \cos (2 \theta ))\right)
    \;.
\end{align}
where we set $\cos(\theta) = \boldsymbol{\mathsf{r}}^{(1)}\cdot \boldsymbol{\mathsf{r}}^{(2)} / (\mathsf{r}^{(1)} \mathsf{r}^{(2)})$ and $\mathsf{r}^{(1,2)} = |\boldsymbol{\mathsf{r}}^{(1,2)}|$. Plugging it into the action, we arrive at
\begin{equation}
\begin{split}
    e^{-\mathcal{I}^{*}} &\sim \int_0^\pi d\theta \sin(\theta) \int_0^1 d\mathsf{r}^{(1)} \int_0^1 d\mathsf{r}^{(2)} P_{\rm stat}(\mathsf{r}^{(1)}) P_{\rm stat}(\mathsf{r}^{(2)}) \frac{\tr{(\varrho^{(1)}\varrho^{(2)}\varrho^{(1)}\varrho^{(2)})}}{\tr(\varrho^{(1)} \varrho^{(2)})}
 \end{split}
\end{equation}
Integrating over the relative angle $\theta$, we arrive at
\begin{equation}
e^{-\mathcal{I}^{*}} \sim 1- \int_0^1 d\mathsf{r}^{(1)} \int_0^1 d\mathsf{r}^{(2)} P_{\rm stat}(\mathsf{r}^{(1)}) P_{\rm stat}(\mathsf{r}^{(2)}) \frac{\left((\mathsf{r}^{(1)})^2-1\right) \left((\mathsf{r}^{(2)})^2-1\right) \arctanh(\mathsf{r}^{(1)} \mathsf{r}^{(2)})}{2 \mathsf{r}^{(1)}\mathsf{r}^{(2)}}
\end{equation}
Finally, following the same steps as in Sec.~\ref{par:selfconst}, the integrations over the radial variables can be performed explicitly in terms of Bessel functions leading to
\begin{equation}
\begin{split}
    e^{-\mathcal{I}^{*}} &\sim 1 + x^2 -x k(x) (1 + x k(x))
\end{split}
\end{equation}
which coincides with the expression in the main text.
For $\gamma \ll 1$, $\mathcal{I}^\ast$ approaches its maximal value $\log 2$: expanding the closed form at large $x^\ast \simeq 3/(8\gamma)$, one finds $\mathcal{I}^\ast = \log 2 - 4 \gamma + O(\gamma^2)$.
For $\gamma \gg 1$ instead
\begin{equation}
    x^\ast \sim 2 e^{-1- \gamma_{\rm EM}} e^{- 2 \gamma} \;,\hspace{1cm}
    \mathcal{I}^\ast \simeq 2 \gamma \,(x^\ast)^2 \sim 8 \gamma\, e^{-2-2 \gamma_{\rm EM} - 4 \gamma}  \approx 0.34\, \gamma\, e^{-4 \gamma} \;,
\end{equation}
where $\gamma_{\rm EM}$ is the Euler-Mascheroni constant.

The same instantaneous-jump construction can be extended, in principle, to higher R\'enyi entropies by replacing the
swap operator with the appropriate cyclic permutation acting on a larger number of replicas. In that case the relevant
boundary conditions connect more general permutation sectors, and the leading slow-purification regime is expected to
depend on the minimal domain-wall structure connecting them. It remains an interesting open question whether this
minimal contribution is always controlled by elementary domain walls, as in the purity case, or whether genuinely more
complicated permutation sectors can dominate for higher R\'enyi indices.

\end{document}